\documentclass[fleqn,10pt]{wlscirep}
\usepackage{lineno}
\usepackage{amsmath}
\usepackage{amsfonts}
\usepackage{color}
\usepackage{comment}
\usepackage{subfigure}
\usepackage{graphicx}
\usepackage{multirow}
\usepackage{siunitx}
\usepackage{cite}
\usepackage{mathtools}
\usepackage{threeparttable}
\usepackage{tablefootnote}
\usepackage{algorithm, algorithmic}
\usepackage{xspace}
\usepackage{amssymb}
\usepackage{booktabs}
\usepackage{bm}
\usepackage{marvosym}
\usepackage{ifsym}
\usepackage{makecell}
\title{An Equivariant Generative Framework for Molecular Graph-Structure Co-Design}

\author[1,2]{Zaixi Zhang}
\author[1,2]{Qi Liu\Letter}
\author[3]{Chee-Kong Lee}
\author[4]{Chang-Yu Hsieh}
\author[1,2]{Enhong Chen}
\affil[1]{Anhui Province Key Lab of Big Data Analysis and Application, University of Science and Technology of China, Hefei, Anhui 230026, China}
\affil[2]{State Key Laboratory of Cognitive Intelligence, Hefei, Anhui, 230088, China}
\affil[3]{Tencent America, Palo Alto, CA 94306, United States}
\affil[4]{Innovation Institute for Artificial Intelligence in Medicine of Zhejiang University, College of Pharmaceutical Sciences, Zhejiang University, Hangzhou, Zhejiang, 310058, China}

\affil[\Letter]{qiliuql@ustc.edu.cn}

\begin{abstract}
Designing molecules with desirable physiochemical properties and functionalities is a long-standing challenge in chemistry, material science, and drug discovery. Recently, machine learning-based generative models have emerged as promising approaches for \emph{de novo} molecule design.
However, further refinement of methodology is highly desired as most existing methods lack unified modeling of 2D topology and 3D geometry information and fail to effectively learn the structure-property relationship for molecule design.
Here we present MolCode, a roto-translation equivariant generative framework for \underline{Mol}ecular graph-structure \underline{Co-de}sign. In MolCode, 3D geometric information empowers the molecular 2D graph generation, which in turn helps guide the prediction of molecular 3D structure. 
Extensive experimental results show that MolCode outperforms previous methods on a series of challenging tasks
including \emph{de novo} molecule design, targeted molecule discovery, and structure-based drug design.
Particularly, MolCode not only consistently generates  valid (99.95$\%$ Validity) and diverse (98.75$\%$ Uniqueness) molecular graphs/structures with desirable properties, but also generate drug-like molecules with high affinity to target proteins (61.8$\%$ high affinity ratio),
which demonstrates MolCode's potential applications in material design and drug discovery. 
Our extensive investigation reveals that the 2D topology and 3D geometry contain intrinsically complementary information in molecule design, and provides new insights into machine learning-based molecule representation and generation.
\end{abstract}
\begin{document}

\flushbottom
\maketitle
%
%
\thispagestyle{empty}

\section*{Introduction}
Designing molecules with desirable characteristics is of fundamental importance in many applications, ranging from drug discovery \cite{hajduk2007decade, lawson2012antibody, wang2022molecular}, catalysis \cite{freeze2019search} to semiconductors \cite{gomez2016design, xu2016recent}. However, the size of the chemical space is estimated to be in the order of $10^{60}$~\cite{polishchuk2013estimation}, which precludes an exhaustive computational or experimental search of possible molecular candidates. In recent years, advances in machine learning (ML) methods have greatly accelerated the exploration of chemical compound space \cite{butler2018machine, vamathevan2019applications, ekins2019exploiting, von2020exploring, westermayr2021perspective, ceriotti2021machine,keith2021combining, fang2022geometry, wang2022efficient, madani2023large, zhang2021graph, zhang2021motif}.
Many studies propose to generate 2D/3D molecules and optimize molecular properties with deep generative models\cite{you2018graph, shi2020graphaf, gebauer2019symmetry, wang2021multi, gebauer2022inverse, zhang2023molecule}.

Molecules can be naturally represented as 2D graphs where nodes denote atoms, and edges represent covalent bonds. Such concise representation has motivated a series of studies in the tasks of molecule design and optimization. These works either predict the atom type and adjacency matrix of the graph\cite{ma2018constrained, de2018molgan,zang2020moflow, madhawa2019graphnvp}, or employ autoregressive models to sequentially add nodes and edges\cite{shi2020graphaf,luo2021graphdf}. Furthermore, some methods leverage the chemical priors of molecular fragments/motifs and propose to generate molecular graphs 
fragment-by-fragment \cite{jin2018junction, jin2020hierarchical}. However, complete information about a molecule cannot be obtained from these methods since the 3D structures of molecules are still unknown, which limits their practical applications. Due to intramolecular interactions or rotations of structural motifs, the same molecular graph can correspond to various spatial conformations with different quantum properties\cite{ganea2021geomol,xu2021end, shi2021learning,liu2021pre, mahmood2021masked}. Therefore, molecular generative models considering 3D geometry information are desired to better learn structure-property relationships.

Recently, some studies characterize molecules as 3D point clouds where each point has atom features (\emph{e.g.,} atom types) and 3D coordinates and corresponding generative models have been proposed for 3D molecule design. These methods include estimating pairwise distances between atoms \cite{hoffmann2019generating}, employing diffusion models to predict atom types and coordinates of all atoms \cite{hoogeboom2022equivariant}, and using autoregressive models to place atoms in 3D space step-by-step \cite{gebauer2019symmetry, luo2021autoregressive, gebauer2022inverse}. Since molecular drugs inhibit or activate particular biological functions by binding to the target proteins, another line of work further proposes generating 3D molecules inside the target protein pocket, which is a complex conditional generation task \cite{luo20213d, mendez2021geometric, liu2022generating, peng2022pocket2mol}.
However, most of these methods do not explicitly consider chemical bonds and valency constraints and may generate molecules that are not chemically valid. Moreover, the lack of bonding information also inhibits the generation of realistic substructures (\emph{e.g.,} benzene rings). 

In this work, we propose MolCode, a roto-translation equivariant generative model for \underline{Mol}ecular graph-structure \underline{Co-de}sign from scratch or conditioned on the target protein pockets. 
Our model is motivated by the intuition that \emph{the information of the 2D graph and 3D structure is intrinsically complementary to each other in molecule generation}: the 3D geometric structure information empowers the generation of chemical bonds, and the bonding information can in turn guide the prediction of 3D coordinates to generate more realistic substructures by constraining the searching space of bond length/angles.
In MolCode, we employ autoregressive flow as the backbone framework to generate atom types, chemical bonds, and 3D coordinates sequentially. To encode intermediate 3D graphs, roto-translation equivariant graph neural networks (GNNs) \cite{liu2021spherical, satorras2021n} are first used to obtain node embeddings. Note that our MolCode is agnostic to the choice of encoding GNNs.
Then, a novel attention mechanism with bond encoding enriches embeddings with global context as well as bonding information.   
In the decoding process, we construct a local
coordinate system based on local reference atoms and predict the relative coordinates, ensuring the equivariance property of atomic coordinates and the invariance property of likelihood. The generated 2D molecular graphs also help check the chemical validity of the generated molecules in each step.
In our experiments, we show that MolCode outperforms existing generative models in generating diverse, valid, and realistic molecular graphs and structures from scratch. Further investigations on targeted molecule discovery show that MolCode can generate molecules with desirable properties that are scarce in the training set, demonstrating its strong capability of capturing structure-property relationships for generalization.
Finally, we extend MolCode to the structure-based drug design task and manage to generate drug-like ligand molecules with high binding affinities.
Systematic hyperparameter analysis and ablation studies show that MolCode is robust to hyperparameters and the unified modeling of 2D topology and 3D geometry consistently improves molecular generation performance. 
\begin{figure}[t]
	\centering
	\includegraphics[width=0.98\linewidth]{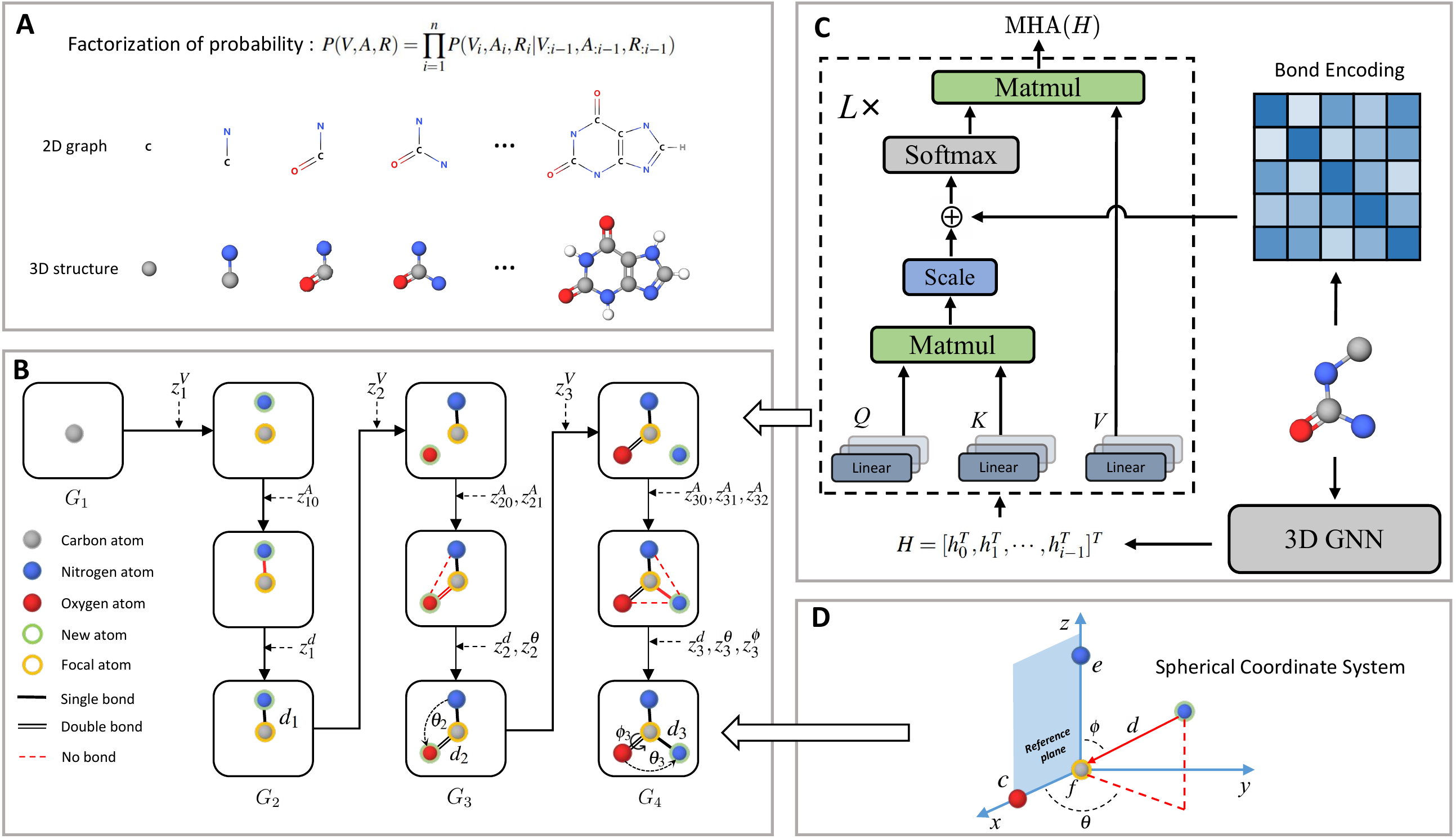}
	\caption{\textbf{Molecule generation with MolCode. a}, In the sequential generation, MolCode concurrently generates molecular 2D graphs and 3D structures. The joint probability of atom types, bond types, and coordinates can then be factorized into a chain of conditional probabilities. {\bf b}, MolCode employs the normalized flow as the backbone model and predicts atom types, bond types, and coordinates sequentially in each step. {\bf c}, MolCode employs roto-translation equivariant Graph Neural Networks and multi-head self-attention with bond encoding for the conditional feature extraction from the intermediate 3D graph. {\bf d}, For the generation of atomic coordinates, MolCode firstly constructs a local spherical coordinate system and generates the relative coordinates i.e. $d, \theta, \phi$, which ensure the equivariance of coordinates and the invariance of likelihood.}
	\label{MolCode}
 \vspace{-1.5em}
\end{figure}

\section*{Results}
\subsection*{Sequential Generation with Flow Models}
Contrary to previous works that treat molecules solely as 2D graphs or 3D point clouds, a molecule is comprehensively represented as a 3D-dimensional graph $G=(V,A,R)$ in this work. Let $a$ and $b$ denote the number of atom types and bond types. For a molecule with $n$ atoms, $V \in \{0,1\}^{n\times a}$ is the atom type matrix, $A \in \{0,1\}^{n\times n \times (b+1)}$ is an adjacency matrix, and $R \in \mathbb{R}^{n\times 3}$ is the 3D atomic coordinate matrix.
We add one additional type of edge
between two atoms, which corresponds to no edge between two atoms. Following previous works like GraphAF\cite{shi2020graphaf} and G-SchNet\cite{gebauer2019symmetry}, we formalize the problem of molecular graph generation as a sequential decision process (Fig. \ref{MolCode}a and b). We can factorize the probability of molecule $P(V,A,R)$ as:

\vspace{-1em}
\begin{align}
    P(V,A,R)&=\prod \limits_{i=1}^n P(V_{i}, A_{i}, R_{i}|V_{:i-1},A_{:i-1},R_{:i-1})\\
    &=\prod \limits_{i=1}^n \prod\limits_{j=0}^{i-1} P(V_{i}|V_{:i-1},A_{:i-1},R_{:i-1})\cdot P(A_{ij}|V_{:i},A_{:i-1},R_{:i-1})\cdot P(R_i|V_{:i},A_{:i},R_{:i-1}),
\end{align}
where $V_{:i-1}, A_{:i-1}$ and $R_{:i-1}$ indicate the graph $(V,A,R)$
restricted to the first $i-1$ atoms, $V_i$ and $R_i$ represent the atom type and coordinates of the $i$-th atom, and $A_i$ denotes the connectivity of the $i$-th atom to the first $i-1$ atoms. We employ a normalized flow model \cite{papamakarios2021normalizing} to learn such probabilities.
A flow model aims to learn a parameterized invertible function between the data point variable $x$ and the latent variable $z$: $f_\theta: z\in \mathbb{R}^d \xrightarrow{} x \in \mathbb{R}^d$.
The latent distribution $p_Z$ is a pre-defined probability distribution, \textit{e.g.,} a Gaussian distribution. The data distribution $p_X$ is unknown. But given a data point $x$, its log-likelihood can be computed with the change-of-variable theorem:
\begin{eqnarray}
    {\rm log} p_X(x)= {\rm log}p_Z(f_\theta^{-1}(x)) + {\rm log}|{\rm det}J|,
    \label{normalized flow}
\end{eqnarray}
where $J=\frac{\partial f_\theta^{-1}(x)}{\partial x}$ denotes the Jacobian matrix. To train the flow model on a molecule dataset, the log-likelihoods of all data points are computed from Eq. (\ref{normalized flow}) and maximized via gradient ascent. In the sampling process, a latent variable $z$ is first sampled from the pre-defined latent distribution $p_Z$. Then the corresponding data point $x$ is obtained by performing the feedforward transformation $x = f_\theta(z)$. Therefore, $f_\theta$ needs to be inevitable, and the computation of ${\rm det}J$ should be tractable for the training and sampling efficiency. A common choice is the affine coupling layers \cite{dinh2014nice, dinh2016density, shi2020graphaf} where the computation of ${\rm det}J$ is very efficient because $J$ is an upper triangular matrix.

Fig. \ref{MolCode} shows a schematic depiction of the MolCode architecture. At each generation step, we predict the new atom type, bond types, and the 3D coordinates sequentially. We use an equivariant graph neural network for the extraction of conditional information from intermediate molecular graphs. A novel multi-head self-attention network with bond encoding is proposed to further capture the global and bonding information. For the generation of atomic coordinates, MolCode firstly constructs a local spherical coordinate system and generates the relative coordinates i.e. $d, \theta, \phi$, which ensure the equivariance of coordinates and the invariance of likelihood. In the \emph{de novo} molecule design and targeted molecule discovery, MolCode generates molecules from scratch. In structure-based drug design, which is a conditional generation task, the target protein pocket represented as a 3D-dimensional graph is first input into MolCode. Then MolCode generates ligand molecules based on the protein pocket.   

We train MolCode on a set of molecular structures and the corresponding molecular graphs can be obtained with toolkits in chemistry\cite{o2011open,kim2015universal}. In the generation process, we check whether the generated bonds violate the valency constraints at each step. If the newly added bond breaks the valency constraint, we just reject it, sample a new latent variable and generate another new bond type. 
More details on the model architecture and training procedure can be found in the Methods section.

\subsection*{\emph{De novo} Molecule Design}
\begin{table}[t]
\caption{\textbf{Results of random molecule generation.} Validity calculates the percentage of valid molecules among all the generated molecules; Uniqueness refers to the percentage of unique molecules among the valid molecules; Novelty measures the fraction of molecules not in the training set among all the valid and unique molecules. 
The best results are bolded.}
\centering

\begin{tabular}{lccc}
\toprule
\textbf{Method}&\textbf{Validity}&\textbf{Uniqueness}& \textbf{Novelty}\\ 
\midrule
E-NFs & 41.30$\%$ &92.96$\%$&81.12$\%$\\ 
G-SchNet & 84.19$\%$&94.11$\%$&\textbf{83.47$\%$}\\
G-SphereNet   & 87.54$\%$&95.49$\%$&81.55$\%$ \\ 
EDM & 92.27$\%$&98.24$\%$& 72.84$\%$\\
MolCode (w/o check) & 94.60$\%$&96.54$\%$ &74.18$\%$\\ 
MolCode (w/o bond)& 92.12$\%$&94.32$\%$&75.43$\%$\\
MolCode      & \textbf{99.95$\%$}&\textbf{98.75$\%$}&75.90$\%$\\ \bottomrule
\label{random results}
\end{tabular}
\vspace{-2em}
\end{table}

\begin{figure*}[t]
	\centering
	\includegraphics[width=0.98\linewidth]{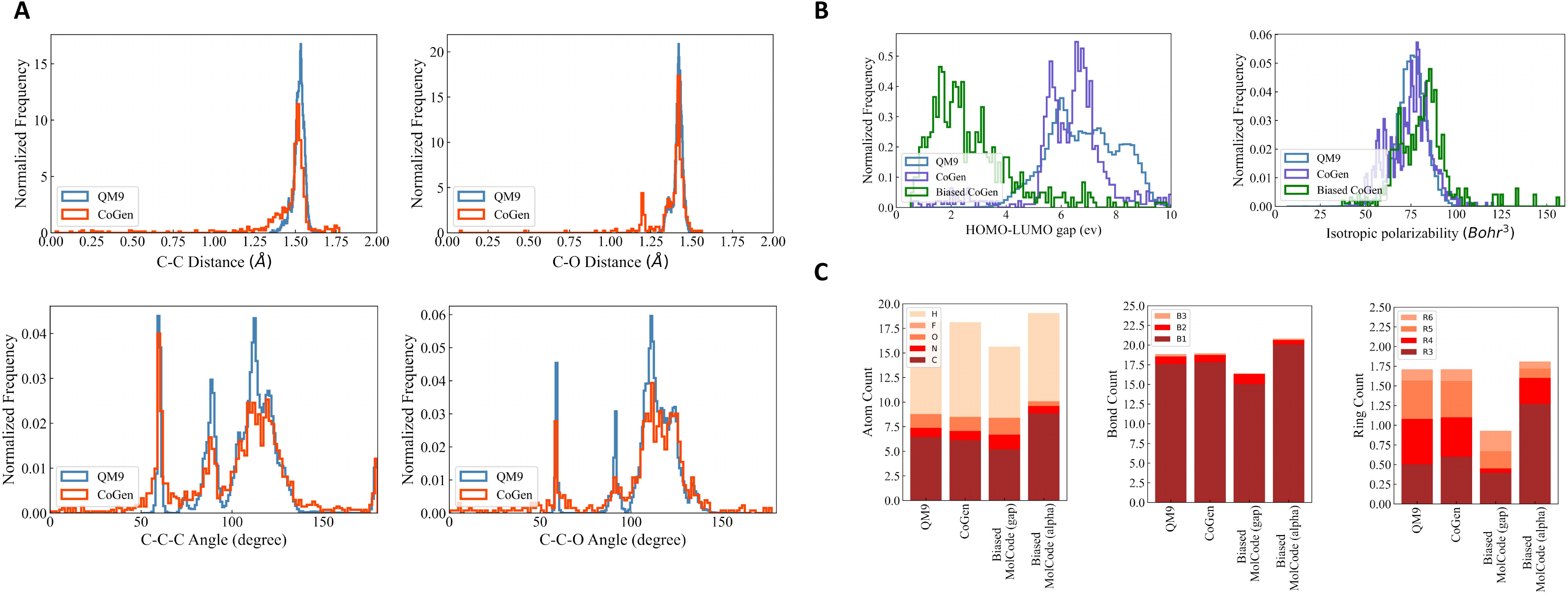}
	\caption{{\bf Results of random molecule generation.} {\bf a}, Radial distribution functions for carbon-carbon single bond and carbon-oxygen single bond (first row) and angular distribution functions for bonded carbon-carbon-carbon and carbon-carbon-oxygen chains (second row) in the training data and in the generated molecules by MolCode. {\bf b}, Histograms of calculated HOMO-LUMO gaps and isotropic polarizability for molecules generated with the biased MolCode (green curves), MolCode before biasing (purple curves), and for the QM9 dataset (blue curves). {\bf c}, Bar plots showing the average numbers of atoms, bonds, and rings per molecule for QM9 and for molecules generated with MolCodes. B1, B2, and B3 correspond to single, double, and triple bonds. R3, R4, R5, and R6 are rings of size 3 to 6.}
	\label{random generation}
 \vspace{-1em}
\end{figure*}

For virtual screening, the generative model should be able to sample a large quantity of valid and diverse molecules from scratch. 
In the random molecule generation task, we evaluate MolCode on the QM9 dataset \cite{ramakrishnan2014quantum} consisting
of ~134k organic molecules with up to nine heavy atoms from
carbon, nitrogen, oxygen, and fluorine. We use Validity, Uniqueness, and Novelty to evaluate the quality of the generated molecules: Validity calculates the percentage of valid molecules among all the generated molecules; Uniqueness is the percentage of unique molecules among all the valid molecules; Novelty measures the fraction of novel molecules among all the valid and unique ones.  Specifically, the 3D molecular structures are first converted to 2D graphs, and the bond types (single, double, triple, or none) are determined based on the distances between pairs of atoms and the atom types \cite{kim2015universal}.
A molecule is considered valid if it obeys the chemical valency rules; it is considered unique or novel if its 2D molecular graph appears only once in the whole sampled molecule set or does not exist in the training set. 
In Table. \ref{random results}, we compare MolCode with four state-of-the-art baselines including E-NFs \cite{garcia2021n}, G-SchNet \cite{gebauer2019symmetry}, G-SphereNet \cite{luo2021autoregressive}, and EDM \cite{hoogeboom2022equivariant} on 3D molecule generation. We also compare MolCode with its two variants i.e. MolCode without validity check (MolCode w/o check) and MolCode without bond information (MolCode w/o bond) for ablation studies. All metrics are computed from 10,000 generated molecular structures. We observe that MolCode achieves the best performance in generating valid and diverse molecular structures (99.95 $\%$ Validity, 98.75 $\%$ Uniqueness). With the advantage of the generated bonds, MolCode can rectify the generation process when the valency constraints are violated, and therefore better explore the chemical space with the autoregressive flow framework. Interestingly, even without a validity check, MolCode can still achieve Validity as high as 94.60 $\%$, which indicates the strong ability of MolCode to capture the underlying chemical rules by modeling the generation of bonds. 
In MolCode (w/o bond), the bonding information is not provided to the conditional information extraction block. The Validity drops from 99.95 $\%$ to 92.12 $\%$ and the Uniqueness drops from 98.75 $\%$ to 94.32 $\%$, which also verifies the usefulness of bonding information in MolCode. Regarding Novelty, as discussed by previous work \cite{hoogeboom2022equivariant} that QM9 is the exhaustive enumeration of molecules that satisfy a predefined set of constraints, the Novelty of MolCode is reasonable and acceptable. 

To further investigate how well our model fits the distribution of QM9, we conduct qualitative substructure analysis (Table. \ref{substructure}). Specifically, we first collect the bond length/angle distributions in the generated molecules and the training dataset and then employ Kullback-Leibler (KL) divergence to compute their distribution distances. We show several common bond and bond angle types. We can observe that MolCode obtains much lower KL divergence than the other methods and its variant without bond information, indicating that the molecules generated by MolCode capture more geometric attributes of data.
Moreover, we show two sets of bond length distributions (carbon-carbon single bond and carbon-oxygen single bond) and two sets of bond angle distributions (carbon-carbon-carbon and carbon-carbon-oxygen chains) in Fig. \ref{random generation}a. Generally, the distributions of MolCode align well with those of QM9, indicating that the distances and angles between atoms are accurately modeled and reproduced.

\begin{table*}[t]
\caption{\textbf{Results of the targeted molecule generation.} We aim to minimize the HOMO-LUMO gap and maximize the isotropic polarizability. The properties are calculated by PySCF and the best results are bolded. \textbf{Good Percentage} measures the ratio of molecules with HOMO-LUMO gaps smaller than 4.5 eV or isotropic polarizabilities larger than 91 Bohr$^3$ respectively. }
\centering
\begin{tabular}{lcccccc}
\toprule
                \multirow{2}{*}{\textbf{Method}}& \multicolumn{3}{c}{\textbf{HOMO-LUMO gap}}&\multicolumn{3}{c}{\textbf{Isotropic polarizability}}  \\ 
                \cmidrule(r){2-4} \cmidrule(r){5-7}&Mean& Optimal& Good Percentage&Mean& Optimal& Good Percentage  \\ \midrule
QM9 (Dataset) &  6.833& 0.669& 3.20$\%$& 75.19& 196.62& 2.04$\%$ \\ 
G-SchNet & 3.332& 0.671 & 75.50$\%$& 78.20 &216.06& 31.39$\%$\\ 
G-SphereNet   & 2.967& 0.315& 81.58$\%$& 87.21& 378.63& 34.72$\%$ \\ 
EDM &3.255&0.453&76.19$\%$&89.10&381.24&33.23$\%$\\
MolCode (w/o check)&2.905&0.284&81.80$\%$&92.20&359.48&36.15$\%$\\
MolCode (w/o bond)& 2.874&0.267&83.56$\%$&90.82&372.19&35.31$\%$\\
MolCode    &\textbf{2.809} &\textbf{0.178}&\textbf{87.76}\% &\textbf{95.36} &\textbf{403.57}&\textbf{38.40}$\%$ \\ 
\bottomrule
\end{tabular}
\label{target generation}
\end{table*}

\subsection*{Targeted Molecule Discovery}
The ability to generate
molecules with desirable properties that are either absent or rare in the
training data (e.g., new materials) is quite useful for the target exploration of chemical space. 
Here we conduct two targeted molecule discovery experiments, namely \emph{minimizing} the HOMO-LUMO gap and \emph{maximizing} the isotropic polarizability.  Following previous works \cite{luo2021autoregressive, gebauer2019symmetry}, we finetune the pretrained generative models on the collected biased datasets. Specifically, we collect all molecular structures whose HOMO-LUMO gaps are smaller than 4.5 eV and all molecular structures whose isotropic polarizabilities are
larger than 91 Bohr$^3$ from the QM9 as the biased datasets. Afterward, we generate 10,000 molecular structures with the finetuned model and compute the quantum properties (HOMO-LUMO gap and isotropic polarizability) with the PySCF package \cite{sun2018pyscf, sun2020recent}.
The performance is then evaluated by calculating the mean and optimal value over all property scores (Mean and Optimal) and the percentage of molecules with good properties (Good Percentage). Molecules with good properties are those with HOMO-LUMO gaps smaller than 4.5 eV and isotropic polarizabilities larger than 91 Bohr$^3$, respectively. 

The results of targeted molecule discovery for two quantum properties are shown in Table. \ref{target generation}.  For both properties, our MolCode outperforms all the baseline methods and its variants without validity check and bonding information, demonstrating
MolCode's strong capability in capturing structure-property relationships and generating molecular structures with desirable properties. For instance, even though the biased datasets are only 3.20$\%$ and 2.04$\%$ of QM9 respectively, the fine-tuned MolCode achieves Good Percentages of 87.76$\%$ and 38.40$\%$. We also illustrate the property distributions of QM9, MolCode, and biased MolCode in Fig. \ref{random generation}b. Clearly, we can observe that the property distributions of MolCode align well with those of the QM9 dataset while the property distributions of the biased MolCodes shift towards smaller HOMO-LUMO gap and larger isotropic polarizability respectively.

Fig. \ref{random generation}c reveals further insights into the structural statistics of the generated molecules. First, we observe that MolCode captures the atom, bond, and ring counts of the QM9 dataset accurately. Second, for the biased MolCode towards smaller HOMO-LUMO gaps, the generated molecules exhibit an
increased number of nitrogen/oxygen atoms and double-bonds in addition to a tendency towards forming six-atom rings. 
These features indicate the presence of aromatic rings with nitrogen/oxygen atoms and conjugated systems with alternating single and double bonds, which are important motifs in organic semiconductors with small HOMO-LUMO gaps. 
Finally, for the biased MolCode towards larger isotropic polarizability, the generated molecules contain more atoms, bonds, and rings, which are the prerequisites for large isotropic polarizabilities.

\subsection*{Structure-based Drug Design}
Designing ligand molecules binding with target proteins is a fundamental and challenging task in drug discovery \cite{anderson2003process}. 
According to the lock
and key model \cite{tripathi2017molecular, alon2021structures}, the molecules that bind tighter to a disease target are more likely to be drug candidates with higher bioactivity against the disease. Therefore, it is beneficial to take the structure of the target proteins into consideration when generating molecules for drug discovery. 
Here, we train MolCode on the CrossDocked2020 dataset \cite{francoeur2020three} which contains 22.5 million protein-molecule complexes for structure-based drug design. Starting with the target protein pocket as the context, MolCode iteratively predicts the ligand atom types, bond types, and atom coordinates. 
We generate 100 ligand molecules for each target protein pocket in the test set. 
More details are included in the Methods section. 

\begin{figure*}[t]
	\centering
	\subfigure{\includegraphics[width=0.3\linewidth]{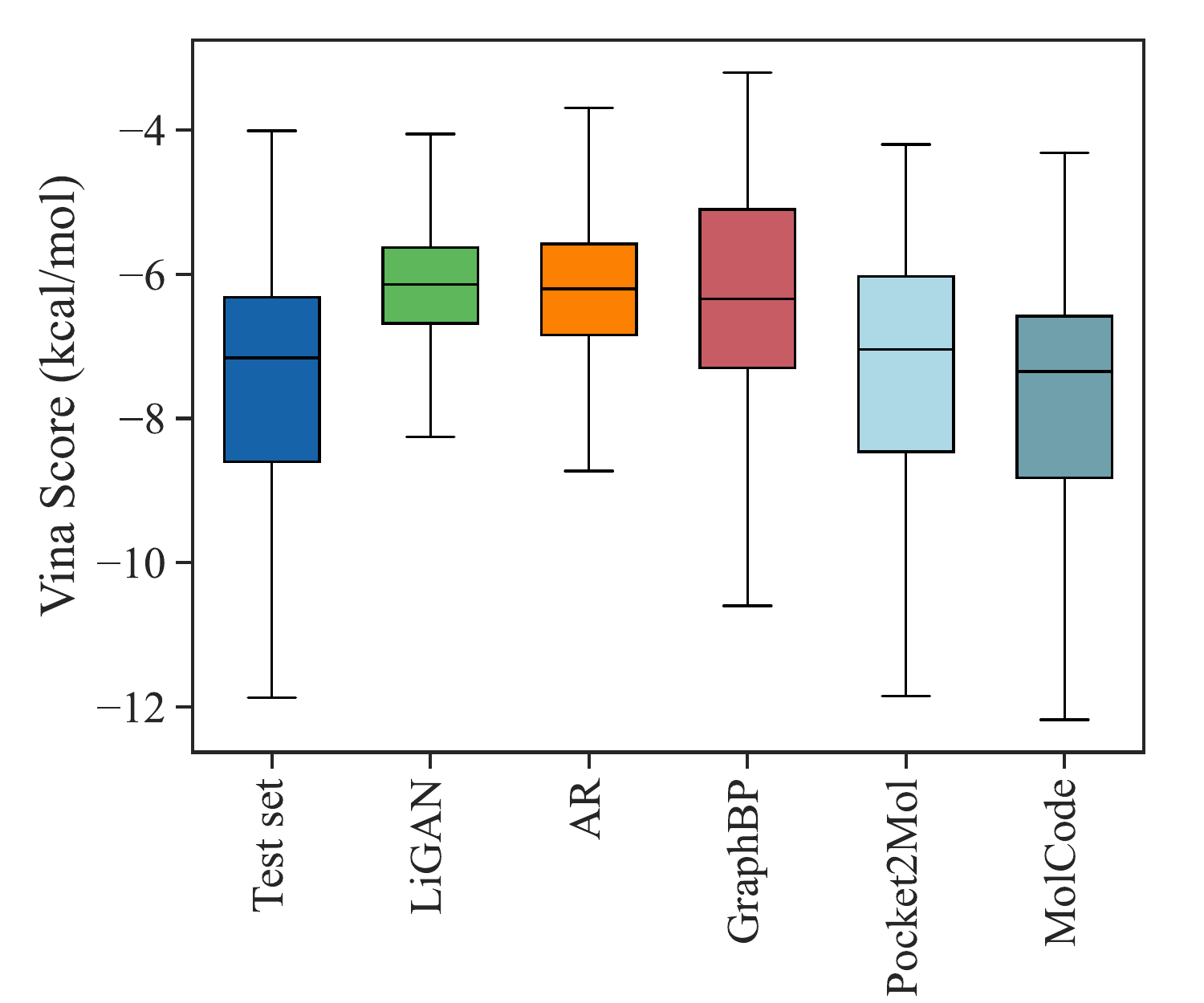}}
    \subfigure{\includegraphics[width=0.3\linewidth]{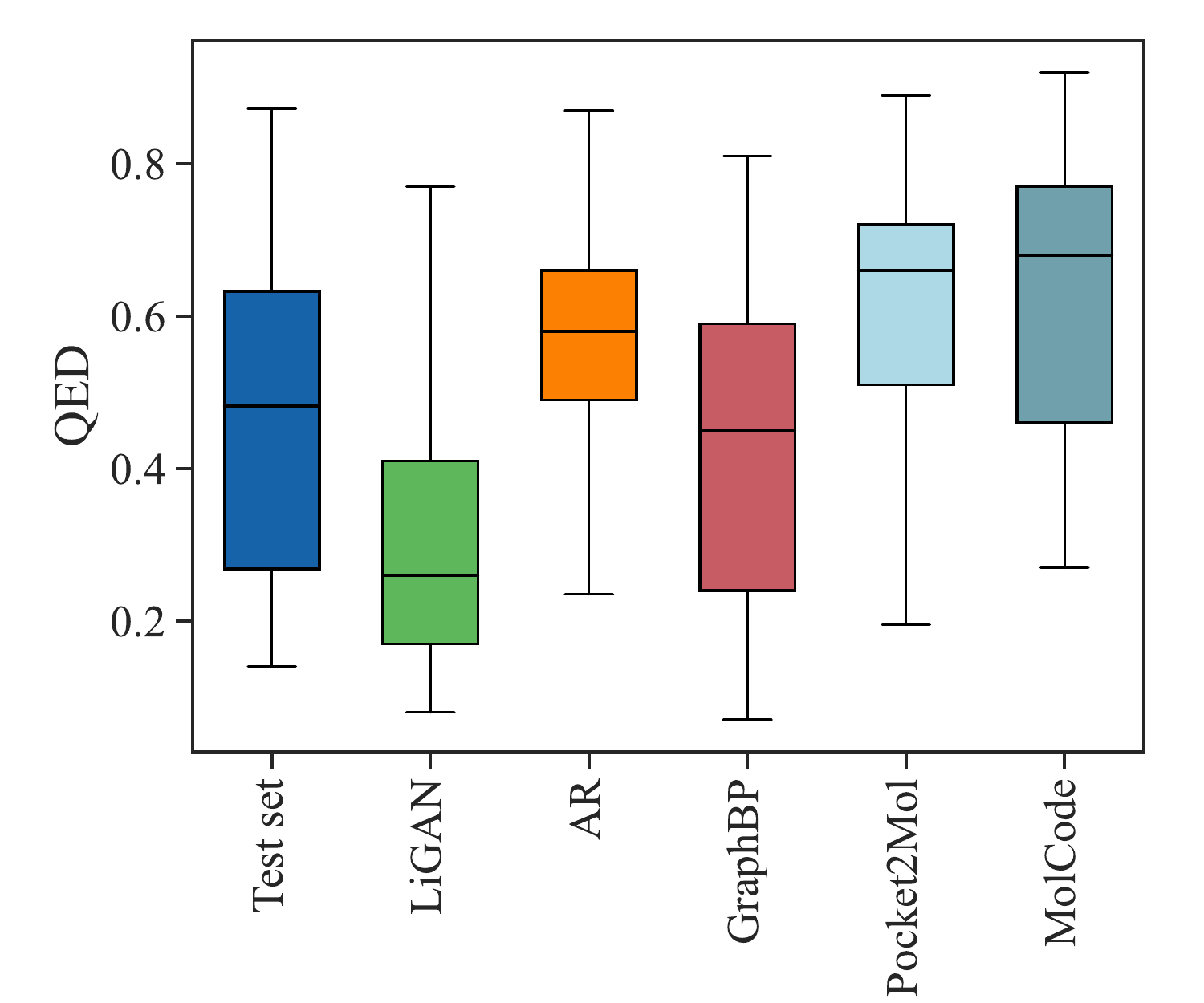}}
    \subfigure{\includegraphics[width=0.3\linewidth]{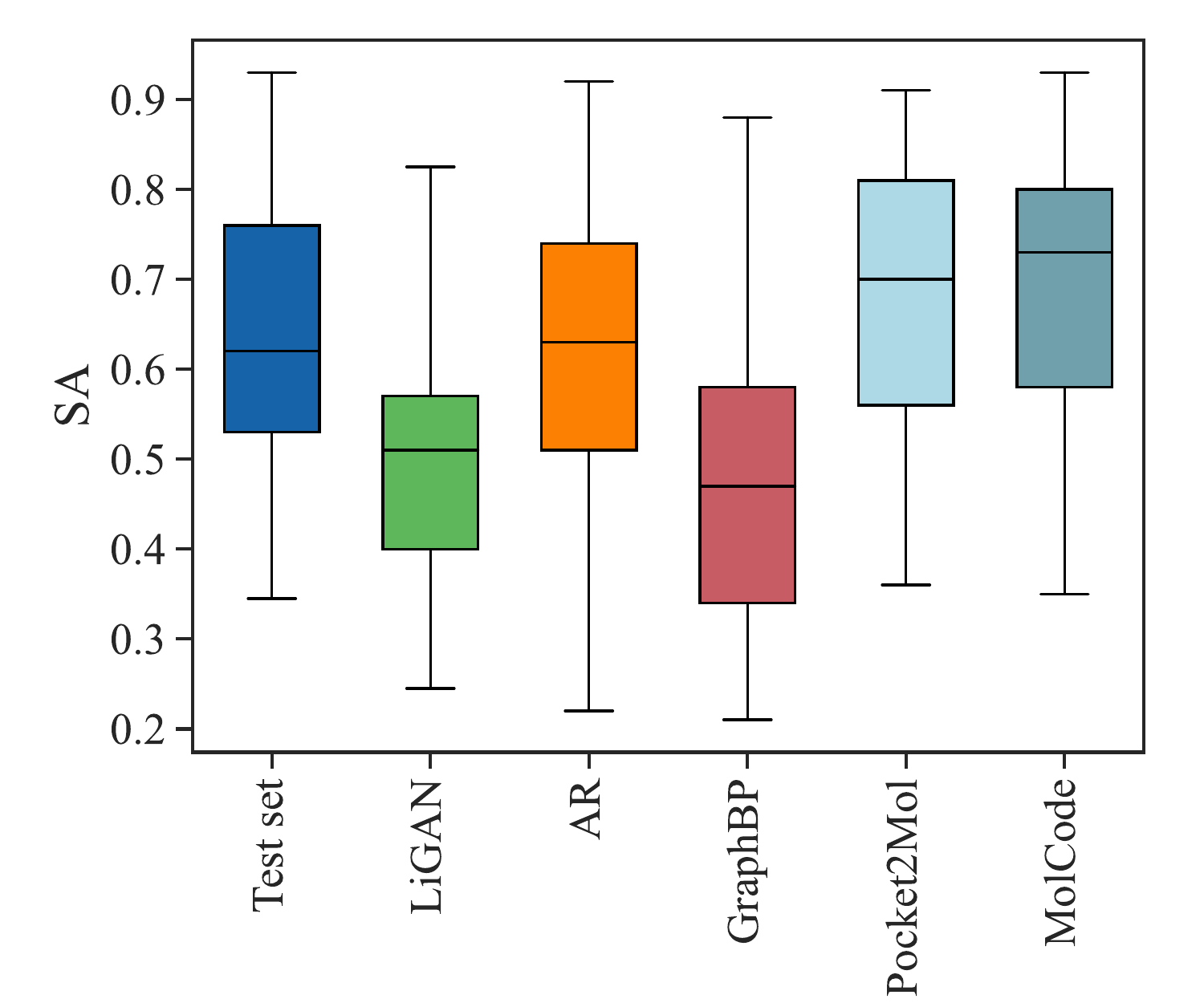}}
	\caption{The distributions of Vina scores, QED, and SA scores of the generated molecules. We also show the distributions of the test set for reference. Lower Vina scores and higher QED and SA indicate better ligand quality.}
	\label{ligand distribution}
\end{figure*}

Fig. \ref{ligand distribution} shows the property distributions of the sampled ligand molecules. Here, we mainly focus on the following metrics following previous works \cite{luo20213d,peng2022pocket2mol}: \textbf{Vina Score} measures the binding affinity between the generated molecules and the protein pockets; \textbf{QED} measures how likely a molecule is a potential drug candidate; \textbf{Synthesizability (SA)} represents the difficulty of drug synthesis (the score is normalized between 0 and 1 and higher values indicate easier synthesis).
In our work, The Vina Score is calculated by QVina \cite{trott2010autodock, alhossary2015fast}, and the chemical properties are calculated by RDKit \cite{bento2020open} over the valid molecules. Before feeding to Vina, all the generated molecular structures are firstly refined by universal force fields \cite{rappe1992uff}.
Four competitive baselines including LiGAN \cite{ragoza2022generating}, AR \cite{luo20213d}, GraphBP \cite{liu2022generating}, and Pocket2Mol \cite{peng2022pocket2mol} are compared. We also show the distributions of the test set for reference. MolCode can generate ligand molecules with higher binding affinities (lower Vina scores) than baseline methods. 
Specifically, MolCode succeeds to generate molecules with higher affinity than corresponding reference molecules for 61.8$\%$ protein pockets on average. 
Moreover, the generated molecules also exhibit more potential to be drug candidates (higher QED and SA). These improvements indicate that MolCode effectively captures the distribution of 3D ligand molecules conditioned on binding sites with the graph-structure co-design scheme.

\begin{figure*}[t]
    \centering
    \includegraphics[width=0.88\linewidth]{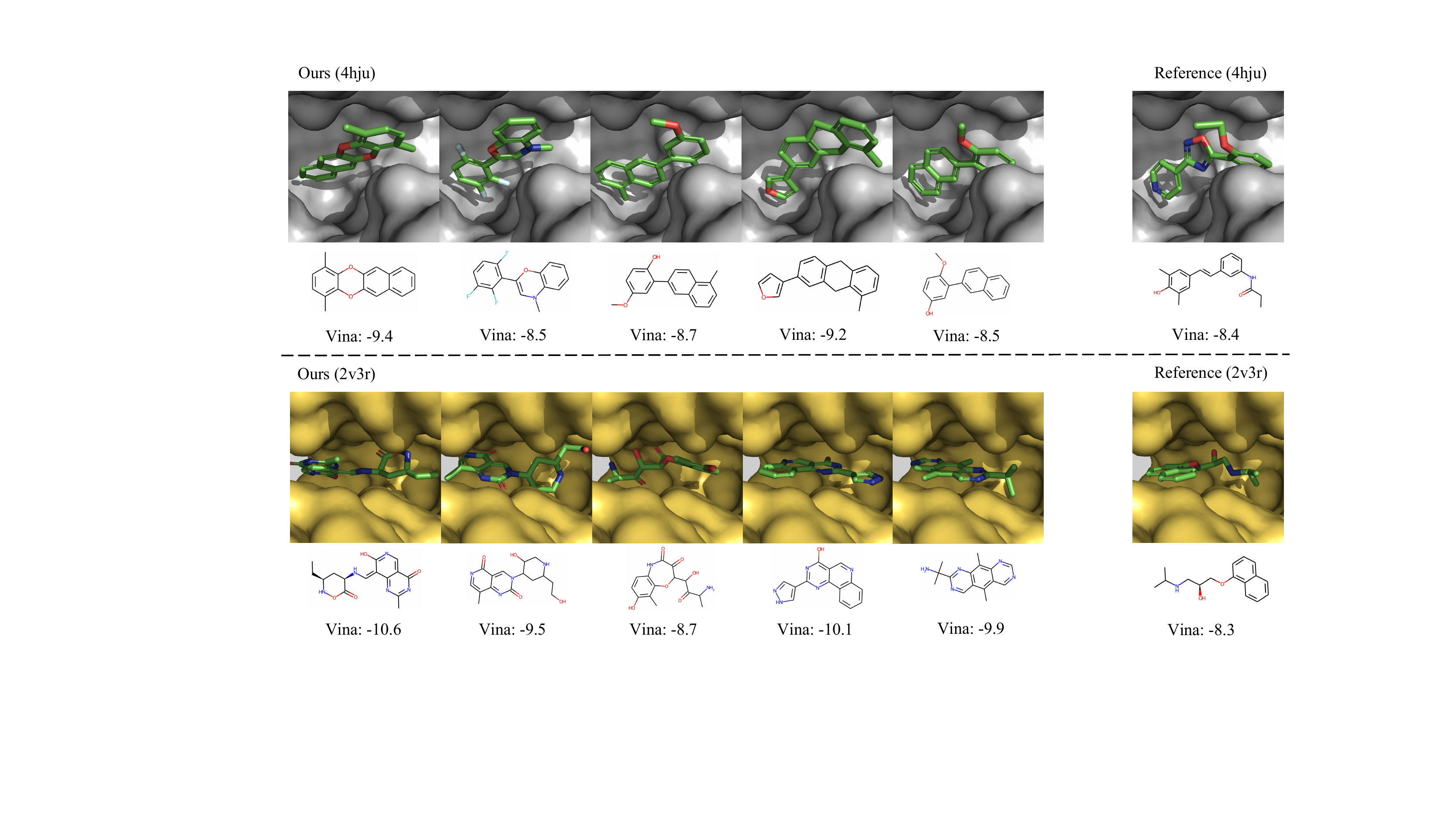}
    \caption{Examples of the generated molecules with higher binding affinities than the references. We report the Vina scores and a lower Vina score indicates higher binding affinity.}
    \label{case}
\end{figure*}

In Fig.\ref{case}, we further show several examples of generated 3D molecules with higher affinities to the target proteins than their corresponding reference molecules in the test set. It can be observed that our generated molecules with higher
binding affinity also have diverse structures and are largely different from the reference molecules. It demonstrates that MolCode is capable
of generating diverse and novel molecules to bind target proteins, instead of just memorizing and reproducing known molecules in the dataset, which is quite important in exploring novel drug candidates.

\section*{Conclusion}
In this article, we have reported a roto-translation equivariant generative framework for molecular graph-structure co-design from scratch or conditioned on the target protein pockets. As compared to existing methods that only represent and generate 2D topology graphs or 3D geometric structures, MolCode concurrently designs 2D molecular graphs and 3D structures and can well capture complex molecular  relationships. Extensive experiments on \emph{de novo} molecule design, targeted molecule discovery, and structure-based drug design demonstrate the effectiveness of our model. Our investigation demonstrates that the 2D topology and 3D geometry contain intrinsically
complementary information for molecular representation and generation and the unified modeling of them greatly improve the molecular generation performance. 

There are also several potential extensions of MolCode as future works. First, MolCode may be extended and applied to significantly larger systems with more diverse atom types such as proteins and crystal materials. Although MolCode has been trained on ligand-protein pocket complexes from the Crossdocked2020 dataset, modifications will be necessary to ensure further scalability and robustness \cite{zhang2022hierarchical, zhang2021graphmi, zhang2022model, zhang2021backdoor, zhang2022protgnn}. 
Another potential improvement is to incorporate chemical priors such as ring structures into MolCode to generate more valid molecules and realistic 3D structures \cite{zhang2021motif, zhang2023molecule}. For example, the molecules may be generated fragment-by-fragment instead of atom-by-atom, which can also speed up the generation process. Furthermore, wet-lab experiments may be conducted to validate the effectiveness of MolCode.
Overall, we anticipate that further developments 
in deep generative models will greatly accelerate and benefit various applications in material design and drug discovery.
\section*{Methods}
\textbf{Dataset.}
For the task of random molecule generation and targeted molecule discovery, we evaluate MolCode on the QM9 \cite{ramakrishnan2014quantum} dataset. The QM9 dataset contains over 134k molecules and their corresponding 3D molecular geometries computed by density functional theory (DFT). In the random molecular geometry generation task, we randomly
select 100k 3D molecular geometries as the training set and 10k 3D molecular geometries as the
validation set. 
For the targeted molecule discovery, we collect all molecular geometries whose HOMO-LUMO gaps are smaller than 4.5 eV and all molecular geometries whose isotropic polarizabilities are
larger than 91 Bohr$^3$ as the finetuning dataset. 

As for the structure-based drug design, we use the CrossDocked dataset \cite{francoeur2020three}
which contains 22.5 million protein-molecule structures following \cite{luo20213d} and \cite{peng2022pocket2mol}.
We filter out data points whose
binding pose RMSD is greater than 1 {\AA} and molecules that can not be sanitized with RDkit \cite{bento2020open}, leading to a refined subset with around 160k data points. 
We use mmseqs2 \cite{steinegger2017mmseqs2} to cluster data at 30$\%$ sequence identity, and randomly draw 100,000
protein-ligand pairs for training and 100 proteins from remaining clusters for testing.
For evaluation, we randomly sample 100 molecules for each protein pocket in the test set.

For all the tasks including random/targeted molecule generation and structure-based drug design, MolCode and all the other baseline methods are trained with the same data split for a fair comparison.

\noindent\textbf{Overview of MolCode. }
Let $a$ be the number of atom types, $b$ be the number of bond types, and $n$ denote the number of atoms in a molecule. We can represent the molecule as a 3D-dimensional graph $G = (V, A, R)$, where $V \in \{0,1\}^{n\times a}$ is the atom type matrix, $A \in \{0,1\}^{n\times n \times (b+1)}$ is an adjacency matrix, and $R \in \mathbb{R}^{n\times 3}$ is the 3D atomic coordinate matrix. Note that we add one additional type of edge
between two atoms, which corresponds to no edge between two atoms. Here, each element $V_i$ in $V$ and $A_{ij}$ in $A$ are one-hot vectors. $V_{iu} = 1$ and $A_{ijv} = 1$ represent that the $i$-th atom has type $u$ and there is a type $v$ bond between the $i$-th and $j$-th atom respectively. The $i$-th row of the coordinate matrix $R_i$ represents the 3D Cartesian coordinate of the $i$-th atom.

We adopt the autoregressive flow framework \cite{papamakarios2021normalizing} to generate the atom type $V_i$ of the new atom, the bond types $A_{ij}$, and the 3D coordinates at each step.
Since both the node type $V_i$ and the edge type $A_{ij}$ are discrete, which do not fit into a flow-based
model, we adopt the dequantization method \cite{you2018graph, shi2020graphaf} to convert them into continuous numbers by adding real-valued noise as:
\begin{eqnarray}
    \Tilde{V_i} =V_i + u,~ u \sim U(0,1)^a;~\Tilde{A}_{ij} = A_{ij} + u, u \sim U(0,1)^{b+1}, ~i\geq 1.
\end{eqnarray}
where $U(0, 1)$ is the uniform distribution over the interval $(0, 1)$. To generate $V_i$ and $A_{ij}$, we first sample the
latent variable $z_i^V \in \mathbb{R}^a$ and $z_{ij}^A \in \mathbb{R}^{b+1}$
from the standard Gaussian distribution $\mathcal{N} (0, 1)$, and then map $z_i^V$ and $z_{ij}^A$
to $\Tilde{V_i}$ and $\Tilde{A}_{ij}$ respectively by the following affine transformation:
\begin{eqnarray}
    \Tilde{V_i} = s_i^V \odot z_i^V + t_i^V;~\Tilde{A}_{ij} = s_{ij}^A \odot z_{ij}^A + t_{ij}^A, ~i\geq 1, 0\le j \le i-1,
    \label{atom, bond factors}
\end{eqnarray}
where $\odot$ denotes the element-wise multiplication. Both the scale factors ($s_i^V$ and $s_{ij}^A$) and shift factors ($t_i^V$ and $t_{ij}^A$) depend on the conditional information extracted from the intermediate 3D graph $G_i$, which we will discuss later.
After obtaining $\Tilde{V_i}$ and $\Tilde{A}_{ij}$, $V_i$ and $A_{ij}$ can be computed by taking
the argmax of $\Tilde{V_i}$ and $\Tilde{A}_{ij}$ i.e., $V_i$ = one-hot(arg max $\Tilde{V_i}$) and $A_{ij}$ = one-hot(arg max $\Tilde{A}_{ij}$).

However, it is non-trivial to generate coordinates that satisfy the equivariance to rigid transformations and the invariance property of likehood. Inspired by G-SchNet \cite{gebauer2019symmetry}, MolGym \cite{simm2020reinforcement}, and G-SphereNet \cite{luo2021autoregressive}, we choose to construct a local spherical coordinate system and generate the distance $d_i$
,the angle $\theta_i$, and the torsion angle $\phi_i$ w.r.t.
the constructed local SCS. Specifically, we first choose a
focal atom among all atoms in $G_i$, which serves as the reference point for the new atom. The new atom is expected to be placed in the local region of the selected focal atom. Assume that the focal node is the $f$-th node in $G_i$. First, the distance $d_i$ from the focal atom to the new atom is generated, i.e., $d_i = \|R_i - R_f\|$. Then, if $i \geq 2$, the angle $\theta_i \in [0, \pi]$ between the lines $(R_f, R_i)$ and $(R_f, R_c)$ is generated, where $c$ is the closest atom to the focal atom in $G_i$. Finally, if $i \geq 3$, the torsion angle $\phi_i \in [-\pi, \pi]$ formed by planes $(R_f, R_c, R_i)$ and $(R_f, R_c, R_e)$ is generated, where $e$ denotes the atom closest to $c$ but different from $f$ in $G_i$. Similar to $\Tilde{V_i}$ and $\Tilde{A}_{ij}$, $d_i, \theta_i, \phi_i$ can be obtained by:
\begin{eqnarray}
    d_i &= s_{i}^d \odot z_{i}^d + t_{i}^d, ~i\geq 1,\\
    \theta_i &= s_{i}^\theta \odot z_{i}^\theta + t_{i}^\theta, ~i\geq 2,\\
    \phi_i &= s_{i}^\phi \odot z_{i}^\phi + t_{i}^\phi, ~i\geq 3
    \label{scale,shift factors},
\end{eqnarray}
where $z_i^d, z_i^\theta, z_i^\phi \in \mathbb{R}$ are latent variables sampled from standard Gaussian distributions and the
scale factors $s^d_i, s^\theta_i, s^\phi_i \in \mathbb{R}$ and the shift factors $t^d_i, t^\theta_i, t^\phi_i \in \mathbb{R}$ are the functions of $G_i$. The coordinate $R_i$ of the new atom
is computed based on the relative coordinates $d_i, \theta_i
, \phi_i$ and the reference atoms $(f, c, e)$, hence satisfying the roto-translation equivariance property. 

\noindent\textbf{Encoder. }
Generating the atom type, covalent bonds, and 3D position at each step requires capturing the conditional information of the intermediate graph $G_i$ with an equivariant encoder. In MolCode, we use SphereNet \cite{liu2021spherical} for the QM9 dataset and EGNN \cite{satorras2021n} for the CrossDocked2020 dataset to obtain the node embeddings. Note that MolCode is agnostic to the choice of equivariant graph neural networks. SphereNet can capture the complete geometric information inside molecular structures including bond length/angles and dihedral angles but can hardly scale to large molecules due to computational complexity. On the contrary, EGNN only encodes pairwise distances between atoms and is more efficient than SphereNet on systems with more atoms \emph{e.g.,} ligand-protein pocket complexes.
For the input graph $G_i$, let the node embedding matrix computed by 3D GNN be $H = [h_{0}^T,h_1^T,\cdots,h_{i-1}^T]^T \in \mathbb{R}^{i\times d}$, where $h_{j}$ is the embedding of the $j$-th atom and $d$ is the dimension of embedding.

To further encode the information of covalent bonds and capture the global information in the molecule graph, we modify the self-attention mechanism \cite{vaswani2017attention} and propose a novel bond encoding. The multi-head self-attention (MHA) with bond encoding is calculated as:

\vspace{-15pt}
\begin{align}
    &{\rm MHA}(H) = {\rm Con}({\rm ATT}^1, \cdots, {\rm ATT}^h)W_O,
    ~{\rm ATT}^k(H) = {\rm softmax} (\mathcal{A}^k) V^k, \\
    &\mathcal{A}_{ij}^k = \frac{(h_i W_Q^k)(h_j W_K^k)^T}{\sqrt{d}}+{\rm Con}({\rm Emb}(A_{ij}), h_i+h_j)W_E^k, ~V^k=HW_V^k, ~1\le k\le K,
\end{align}
where Con(·) denotes the concatenation operation, ${\rm Emb}(A_{ij})$ is the embedding of the bond between the $i$-th and $j$-th atom, $K$ is number of attention heads, $W_h^Q, W_h^K, W_h^V, W_h^E,$ and $W^O$ are learnable matrices. 

In MolCode, we use the SphereNet
\cite{liu2021spherical} with 4 layers or EGNN\cite{satorras2021n} with 6 layers to extract features from the intermediate 3D graphs, where the input
embedding size is set to 64 and the output embedding size is set to 256. 
The cutoff is set as 5 {\AA}.
The node features are initialized to the one-hot vectors of atom types and the edge
features are initialized by spherical basis functions. In the multi-head self-attention module with bond encoding, there are 4 attention heads. In addition, we employ 6 flow
layers with a hidden dimension of 128 for the decoder. We use the model configuration for all the experiments.

\noindent\textbf{Decoder. }
To generate new atoms, the scale factor $s_i^V$ and shift factor $t_i^V$ in Eq. (\ref{atom, bond factors}) can be computed as:

\vspace{-15pt}
\begin{align}
    s_i^V, t_i^V = {\rm MLP}({\rm Con}(h_f, {\rm MHA^V}(H)_f)),
\end{align}
where ${\rm MLP}^V$ is a multi-layer perceptron and ${\rm MHA}^V(H)_f$ denotes the $f$-th node embedding from the output of the multi-head self-attention network. With the predicted new atom $V_i$, we can update $H$ to $\Tilde{H}$ and predict $s_{ij}^A$ and $t_{ij}^A$ in Eq. (\ref{atom, bond factors}):

\vspace{-15pt}
\begin{align}
&\Tilde{H} = [h_{0}^T,h_1^T,\cdots,h_{i-1}^T, \Tilde{h}_i^T]^T, ~\Tilde{h}_i={\rm Emb}(V_i),\\
&s_{ij}^A, t_{ij}^A = {\rm MLP}^A({\rm Con}( \Tilde{h}_i, h_j, {\rm MHA}(\Tilde{H})_f)), 0\le j \le i-1,
\end{align}
where ${\rm Emb}(V_i)$ denotes the atom type embedding here.
As for the scale and shift factors in Eq. (\ref{scale,shift factors}), we have: 

\vspace{-15pt}
\begin{align}
    s_{i}^d, t_{i}^d &= {\rm MLP}^d({\rm Con}(h_f, {\rm MHA}(\Tilde{H})_i)),~i \geq 1,\\
    s_{i}^\theta, t_{i}^\theta &= {\rm MLP}^\theta({\rm Con}(h_f, h_c, {\rm MHA}(\Tilde{H})_i)),~i \geq 2,\\
    s_{i}^\phi, t_{i}^\phi &= {\rm MLP}^\phi({\rm Con}(h_f, h_c, h_e, {\rm MHA}(\Tilde{H})_i)),~i \geq 3,
\end{align}
where ${\rm MHA}(\Tilde{H})_i$ is the node embedding of the newly added atom from the output of the multi-head self-attention network.

As for the focal atom selection, we employ a multi-layer perceptron (MLP) with
the atom embeddings as input. 
Atoms that are not valence filled are labeled 1, otherwise 0. 
Particularly, in the structure-based drug design task where there is no ligand atom at the beginning, the focal atoms are defined as protein
atoms that have ground-truth ligand atoms within 4 {\AA} at the first step. After the generation of the first ligand atom, MolCode selects focal atoms from the generated ligand atoms.
At the inference stage,
we randomly choose the focal atom $f$ from atoms whose classification scores are higher than 0.5. The sequential generation process stops if all the classification scores are lower than 0.5 or there is no generated bond between the newly added atom and the previously generated atoms.

\noindent\textbf{Validity Filter. }
The graph-structure codesign scheme in MolCode makes it feasible to check the chemical validity based on the generated 2D graphs at each step. Specifically, we explicitly consider the valency constraints during sampling to check whether current bonds have exceeded the allowed valency. The valency constraint is defined as:
\begin{eqnarray}
    \sum_j |A_{ij}| \le {\rm Valency}(V_i) {~\rm and~} \sum_i |A_{ij}| \le {\rm Valency}(V_j), 
\end{eqnarray}
where $|A_{ij}|$ denote the order of the chemical bond $A_{ij}$.
If the newly added bond breaks the valency constraint, we will reject the bond $A_{ij}$, sample a new $z_{ij}$ in the latent space and generate another new bond type.

\noindent\textbf{Model Training and Inference. }
To make sure the generated atoms are in the local region of their corresponding reference atoms,
we propose to use Prim’s algorithm to obtain the generation orders of atoms. The weights of the edges are set as the distances between atoms. The first atoms of molecules are randomly sampled in each epoch to encourage the generalization ability of the model. With such obtained trajectories,
MolCode is trained by stochastic gradient descent with the
following loss function.
For a 3D molecular graph $G$ with $n$ atoms $(n > 3)$, we maximize its log-likelihood in Eq. (\ref{log likelihood1}) and (\ref{log likelihood2}) to train the MolCode model.
Besides, the atom-wise classifier for focal atom selection is trained with a binary cross entropy loss. 

\vspace{-10pt}
\begin{align}
    {\rm log} p(G) &= \sum_{i=1}^{n-1}\left[{\rm log} p_{Z_V}(z_i^V)+{\rm log}|\frac{\partial \Tilde{V}_i}{\partial z_i^V}|\right] + \sum_{i=1}^{n-1}\sum_{j=0}^{i-1}\left[{\rm log} p_{Z_A}(z_{ij}^A)+{\rm log}|\frac{\partial \Tilde{A}_{ij}}{\partial z_{ij}^A}|\right]\label{log likelihood1}\\
    & + \sum_{i=1}^{n-1}\left[{\rm log} p_{Z_d}(z_i^d)+{\rm log}|\frac{\partial d_i}{\partial z_i^d}|\right] + \sum_{i=2}^{n-1}\left[{\rm log} p_{Z_\theta}(z_i^\theta)+{\rm log}|\frac{\partial \theta_i}{\partial z_i^\theta}|\right]
    + \sum_{i=3}^{n-1}\left[{\rm log} p_{Z_\phi}(z_i^\phi)+{\rm log}|\frac{\partial \phi_i}{\partial z_i^\phi}|\right].
    \label{log likelihood2}
\end{align}

In the random molecule generation task, our MolCode model is trained with Adam \cite{kingma2014adam} optimizer for 100 epochs, where the learning rate is 0.0001 and the batch size is 64. We report the results corresponding to the epoch with the best validation loss. It takes around 36 hours to train a MolCode from scratch on 1 Tesla V100 GPU.
In the targeted molecule discovery task, the model is fine-tuned with a learning rate of 0.0001 and a batch size of 32. The number of training epochs is 40 for the HOMO-LUMO gap and 80 for
the isotropic polarizability. In the task of structure-based drug design, we train MolCode with Adam optimizer for 100 epochs with a learning rate of 0.0001 and a batch size of 8. $\beta_1$ and $\beta_2$ in Adam is set to 0.9 and 0.999, respectively.
For all the tasks including random/targeted molecule generation and structure-based drug design, MolCode and all the other baseline methods are trained with the same data split for a fair comparison. We run the code provided by the authors to obtain the results of baseline methods.

During generation, we use temperature hyperparameters in the prior Gaussian distributions. Specifically, we change the standard deviation of the Gaussian distribution to the
temperature hyperparameters.
To decide the specific values of temperature hyperparameters, we perform a grid search over $\{0.3, 0.5, 0.7\}$ based on Validity and Uniqueness in random molecule generation to encourage generating more valid and diverse molecules. We use 0.5 for sampling $z_i^V$, 0.5 for sampling $z_i^A$, 0.3 for sampling $z_i^d$, 0.3 for sampling $z_i^\theta$, and 0.7 for sampling $z_i^\phi$ as the default setting. We have the following interesting insights for choosing temperature hyperparameters: To generate valid and diverse molecules, the hidden variables for bond lengths/angles ($z_i^d$ and $z_i^\theta$) are assigned with small temperature hyperparameters (low variance) since the values of a certain type of bond lengths/angles are largely fixed. On the contrary, the torsion angles are more flexible in molecules so that the temperature hyperparameter of $z_i^\phi$ is larger. We use the same fixed temperature hyperparameters for the targeted molecule discovery and structure-based drug design experiments. In Fig. \ref{app:hyper}, we show the hyperparameter analysis with respect to $z_i^V$, $z_i^A$, $z_i^d$, $z_i^\theta$, and $z_i^{\phi}$. The default values with these hyperparameters are set to 0.5. MolCode is generally robust to the choice of hyperparameters and can further benefit from setting appropriate hyperparameter values. 

Algorithm \ref{alg:train} and \ref{alg:generate} show the pseudo-codes of the training and generation process of MolCode for random/targeted molecule generation. Note that to scale to large molecules in experiments, the bonds are only generated and predicted between new atoms and the reference atoms. The pseudo-codes of MolCode for structure-based drug design are similar to Algorithm \ref{alg:train} and \ref{alg:generate}, except that the ligand atoms are generated conditioned on the protein pocket instead of generated from scratch.  

\section*{Data availability}
The data necessary to reproduce our numerical benchmark results are publicly available at https://github.com/divelab/DIG and https://github.com/gnina/models.
\section*{Code availability}
The code used in the study is publicly available from the GitHub repository: https://github.com/zaixizhang/MolCode.

\bibliography{main}

\begin{thebibliography}{10}
\expandafter\ifx\csname url\endcsname\relax
  \def\url#1{\texttt{#1}}\fi
\expandafter\ifx\csname urlprefix\endcsname\relax\def\urlprefix{URL }\fi
\providecommand{\bibinfo}[2]{#2}
\providecommand{\eprint}[2][]{\url{#2}}

\bibitem{hajduk2007decade}
\bibinfo{author}{Hajduk, P.~J.} \& \bibinfo{author}{Greer, J.}
\newblock \bibinfo{title}{A decade of fragment-based drug design: strategic
  advances and lessons learned}.
\newblock \emph{\bibinfo{journal}{Nature reviews Drug discovery}}
  \textbf{\bibinfo{volume}{6}}, \bibinfo{pages}{211--219}
  (\bibinfo{year}{2007}).

\bibitem{lawson2012antibody}
\bibinfo{author}{Lawson, A.~D.}
\newblock \bibinfo{title}{Antibody-enabled small-molecule drug discovery}.
\newblock \emph{\bibinfo{journal}{Nature Reviews Drug Discovery}}
  \textbf{\bibinfo{volume}{11}}, \bibinfo{pages}{519--525}
  (\bibinfo{year}{2012}).

\bibitem{wang2022molecular}
\bibinfo{author}{Wang, Y.}, \bibinfo{author}{Wang, J.}, \bibinfo{author}{Cao,
  Z.} \& \bibinfo{author}{Barati~Farimani, A.}
\newblock \bibinfo{title}{Molecular contrastive learning of representations via
  graph neural networks}.
\newblock \emph{\bibinfo{journal}{Nature Machine Intelligence}}
  \textbf{\bibinfo{volume}{4}}, \bibinfo{pages}{279--287}
  (\bibinfo{year}{2022}).

\bibitem{freeze2019search}
\bibinfo{author}{Freeze, J.~G.}, \bibinfo{author}{Kelly, H.~R.} \&
  \bibinfo{author}{Batista, V.~S.}
\newblock \bibinfo{title}{Search for catalysts by inverse design: artificial
  intelligence, mountain climbers, and alchemists}.
\newblock \emph{\bibinfo{journal}{Chemical reviews}}
  \textbf{\bibinfo{volume}{119}}, \bibinfo{pages}{6595--6612}
  (\bibinfo{year}{2019}).

\bibitem{gomez2016design}
\bibinfo{author}{G{\'o}mez-Bombarelli, R.} \emph{et~al.}
\newblock \bibinfo{title}{Design of efficient molecular organic light-emitting
  diodes by a high-throughput virtual screening and experimental approach}.
\newblock \emph{\bibinfo{journal}{Nature materials}}
  \textbf{\bibinfo{volume}{15}}, \bibinfo{pages}{1120--1127}
  (\bibinfo{year}{2016}).

\bibitem{xu2016recent}
\bibinfo{author}{Xu, R.-P.}, \bibinfo{author}{Li, Y.-Q.} \&
  \bibinfo{author}{Tang, J.-X.}
\newblock \bibinfo{title}{Recent advances in flexible organic light-emitting
  diodes}.
\newblock \emph{\bibinfo{journal}{Journal of Materials Chemistry C}}
  \textbf{\bibinfo{volume}{4}}, \bibinfo{pages}{9116--9142}
  (\bibinfo{year}{2016}).

\bibitem{polishchuk2013estimation}
\bibinfo{author}{Polishchuk, P.~G.}, \bibinfo{author}{Madzhidov, T.~I.} \&
  \bibinfo{author}{Varnek, A.}
\newblock \bibinfo{title}{Estimation of the size of drug-like chemical space
  based on gdb-17 data}.
\newblock \emph{\bibinfo{journal}{Journal of computer-aided molecular design}}
  \textbf{\bibinfo{volume}{27}}, \bibinfo{pages}{675--679}
  (\bibinfo{year}{2013}).

\bibitem{butler2018machine}
\bibinfo{author}{Butler, K.~T.}, \bibinfo{author}{Davies, D.~W.},
  \bibinfo{author}{Cartwright, H.}, \bibinfo{author}{Isayev, O.} \&
  \bibinfo{author}{Walsh, A.}
\newblock \bibinfo{title}{Machine learning for molecular and materials
  science}.
\newblock \emph{\bibinfo{journal}{Nature}} \textbf{\bibinfo{volume}{559}},
  \bibinfo{pages}{547--555} (\bibinfo{year}{2018}).

\bibitem{vamathevan2019applications}
\bibinfo{author}{Vamathevan, J.} \emph{et~al.}
\newblock \bibinfo{title}{Applications of machine learning in drug discovery
  and development}.
\newblock \emph{\bibinfo{journal}{Nature reviews Drug discovery}}
  \textbf{\bibinfo{volume}{18}}, \bibinfo{pages}{463--477}
  (\bibinfo{year}{2019}).

\bibitem{ekins2019exploiting}
\bibinfo{author}{Ekins, S.} \emph{et~al.}
\newblock \bibinfo{title}{Exploiting machine learning for end-to-end drug
  discovery and development}.
\newblock \emph{\bibinfo{journal}{Nature materials}}
  \textbf{\bibinfo{volume}{18}}, \bibinfo{pages}{435--441}
  (\bibinfo{year}{2019}).

\bibitem{von2020exploring}
\bibinfo{author}{von Lilienfeld, O.~A.}, \bibinfo{author}{M{\"u}ller, K.-R.} \&
  \bibinfo{author}{Tkatchenko, A.}
\newblock \bibinfo{title}{Exploring chemical compound space with quantum-based
  machine learning}.
\newblock \emph{\bibinfo{journal}{Nature Reviews Chemistry}}
  \textbf{\bibinfo{volume}{4}}, \bibinfo{pages}{347--358}
  (\bibinfo{year}{2020}).

\bibitem{westermayr2021perspective}
\bibinfo{author}{Westermayr, J.}, \bibinfo{author}{Gastegger, M.},
  \bibinfo{author}{Sch{\"u}tt, K.~T.} \& \bibinfo{author}{Maurer, R.~J.}
\newblock \bibinfo{title}{Perspective on integrating machine learning into
  computational chemistry and materials science}.
\newblock \emph{\bibinfo{journal}{The Journal of Chemical Physics}}
  \textbf{\bibinfo{volume}{154}}, \bibinfo{pages}{230903}
  (\bibinfo{year}{2021}).

\bibitem{ceriotti2021machine}
\bibinfo{author}{Ceriotti, M.}, \bibinfo{author}{Clementi, C.} \&
  \bibinfo{author}{Anatole~von Lilienfeld, O.}
\newblock \bibinfo{title}{Machine learning meets chemical physics}
  (\bibinfo{year}{2021}).

\bibitem{keith2021combining}
\bibinfo{author}{Keith, J.~A.} \emph{et~al.}
\newblock \bibinfo{title}{Combining machine learning and computational
  chemistry for predictive insights into chemical systems}.
\newblock \emph{\bibinfo{journal}{Chemical reviews}}
  \textbf{\bibinfo{volume}{121}}, \bibinfo{pages}{9816--9872}
  (\bibinfo{year}{2021}).

\bibitem{fang2022geometry}
\bibinfo{author}{Fang, X.} \emph{et~al.}
\newblock \bibinfo{title}{Geometry-enhanced molecular representation learning
  for property prediction}.
\newblock \emph{\bibinfo{journal}{Nature Machine Intelligence}}
  \textbf{\bibinfo{volume}{4}}, \bibinfo{pages}{127--134}
  (\bibinfo{year}{2022}).

\bibitem{wang2022efficient}
\bibinfo{author}{Wang, D.} \emph{et~al.}
\newblock \bibinfo{title}{Efficient sampling of high-dimensional free energy
  landscapes using adaptive reinforced dynamics}.
\newblock \emph{\bibinfo{journal}{Nature Computational Science}}
  \textbf{\bibinfo{volume}{2}}, \bibinfo{pages}{20--29} (\bibinfo{year}{2022}).

\bibitem{madani2023large}
\bibinfo{author}{Madani, A.} \emph{et~al.}
\newblock \bibinfo{title}{Large language models generate functional protein
  sequences across diverse families}.
\newblock \emph{\bibinfo{journal}{Nature Biotechnology}} \bibinfo{pages}{1--8}
  (\bibinfo{year}{2023}).

\bibitem{zhang2021graph}
\bibinfo{author}{Zhang, Z.} \emph{et~al.}
\newblock \bibinfo{title}{Graph self-supervised learning for optoelectronic
  properties of organic semiconductors}.
\newblock \emph{\bibinfo{journal}{ICML AI4Science workshop}}
  (\bibinfo{year}{2022}).

\bibitem{zhang2021motif}
\bibinfo{author}{Zhang, Z.}, \bibinfo{author}{Liu, Q.}, \bibinfo{author}{Wang,
  H.}, \bibinfo{author}{Lu, C.} \& \bibinfo{author}{Lee, C.-K.}
\newblock \bibinfo{title}{Motif-based graph self-supervised learning for
  molecular property prediction}.
\newblock \emph{\bibinfo{journal}{Advances in Neural Information Processing
  Systems}} \textbf{\bibinfo{volume}{34}}, \bibinfo{pages}{15870--15882}
  (\bibinfo{year}{2021}).

\bibitem{you2018graph}
\bibinfo{author}{You, J.}, \bibinfo{author}{Liu, B.}, \bibinfo{author}{Ying,
  Z.}, \bibinfo{author}{Pande, V.} \& \bibinfo{author}{Leskovec, J.}
\newblock \bibinfo{title}{Graph convolutional policy network for goal-directed
  molecular graph generation}.
\newblock In \emph{\bibinfo{booktitle}{Advances in neural information
  processing systems}}, \bibinfo{pages}{6410--6421} (\bibinfo{year}{2018}).

\bibitem{shi2020graphaf}
\bibinfo{author}{Shi, C.} \emph{et~al.}
\newblock \bibinfo{title}{Graphaf: a flow-based autoregressive model for
  molecular graph generation}.
\newblock \emph{\bibinfo{journal}{International Conference on Learning
  Representations}}  (\bibinfo{year}{2020}).

\bibitem{gebauer2019symmetry}
\bibinfo{author}{Gebauer, N.}, \bibinfo{author}{Gastegger, M.} \&
  \bibinfo{author}{Sch{\"u}tt, K.}
\newblock \bibinfo{title}{Symmetry-adapted generation of 3d point sets for the
  targeted discovery of molecules}.
\newblock In \emph{\bibinfo{booktitle}{Advances in Neural Information
  Processing Systems}}, \bibinfo{pages}{7566--7578} (\bibinfo{year}{2019}).

\bibitem{wang2021multi}
\bibinfo{author}{Wang, J.} \emph{et~al.}
\newblock \bibinfo{title}{Multi-constraint molecular generation based on
  conditional transformer, knowledge distillation and reinforcement learning}.
\newblock \emph{\bibinfo{journal}{Nature Machine Intelligence}}
  \textbf{\bibinfo{volume}{3}}, \bibinfo{pages}{914--922}
  (\bibinfo{year}{2021}).

\bibitem{gebauer2022inverse}
\bibinfo{author}{Gebauer, N.~W.}, \bibinfo{author}{Gastegger, M.},
  \bibinfo{author}{Hessmann, S.~S.}, \bibinfo{author}{M{\"u}ller, K.-R.} \&
  \bibinfo{author}{Sch{\"u}tt, K.~T.}
\newblock \bibinfo{title}{Inverse design of 3d molecular structures with
  conditional generative neural networks}.
\newblock \emph{\bibinfo{journal}{Nature communications}}
  \textbf{\bibinfo{volume}{13}}, \bibinfo{pages}{1--11} (\bibinfo{year}{2022}).

\bibitem{zhang2023molecule}
\bibinfo{author}{ZHANG, Z.}, \bibinfo{author}{Liu, Q.}, \bibinfo{author}{Zheng,
  S.} \& \bibinfo{author}{Min, Y.}
\newblock \bibinfo{title}{Molecule generation for target protein binding with
  structural motifs}.
\newblock In \emph{\bibinfo{booktitle}{International Conference on Learning
  Representations}} (\bibinfo{year}{2023}).

\bibitem{ma2018constrained}
\bibinfo{author}{Ma, T.}, \bibinfo{author}{Chen, J.} \& \bibinfo{author}{Xiao,
  C.}
\newblock \bibinfo{title}{Constrained generation of semantically valid graphs
  via regularizing variational autoencoders}.
\newblock \emph{\bibinfo{journal}{Advances in Neural Information Processing
  Systems}} \textbf{\bibinfo{volume}{31}} (\bibinfo{year}{2018}).

\bibitem{de2018molgan}
\bibinfo{author}{De~Cao, N.} \& \bibinfo{author}{Kipf, T.}
\newblock \bibinfo{title}{Molgan: An implicit generative model for small
  molecular graphs}.
\newblock \emph{\bibinfo{journal}{ICML 2018 workshop on Theoretical Foundations
  and Applications of Deep Generative Models}}  (\bibinfo{year}{2018}).

\bibitem{zang2020moflow}
\bibinfo{author}{Zang, C.} \& \bibinfo{author}{Wang, F.}
\newblock \bibinfo{title}{Moflow: an invertible flow model for generating
  molecular graphs}.
\newblock In \emph{\bibinfo{booktitle}{Proceedings of the 26th ACM SIGKDD
  International Conference on Knowledge Discovery \& Data Mining}},
  \bibinfo{pages}{617--626} (\bibinfo{year}{2020}).

\bibitem{madhawa2019graphnvp}
\bibinfo{author}{Madhawa, K.}, \bibinfo{author}{Ishiguro, K.},
  \bibinfo{author}{Nakago, K.} \& \bibinfo{author}{Abe, M.}
\newblock \bibinfo{title}{Graphnvp: An invertible flow model for generating
  molecular graphs}.
\newblock \emph{\bibinfo{journal}{arXiv preprint arXiv:1905.11600}}
  (\bibinfo{year}{2019}).

\bibitem{luo2021graphdf}
\bibinfo{author}{Luo, Y.}, \bibinfo{author}{Yan, K.} \& \bibinfo{author}{Ji,
  S.}
\newblock \bibinfo{title}{Graphdf: A discrete flow model for molecular graph
  generation}.
\newblock In \emph{\bibinfo{booktitle}{International Conference on Machine
  Learning}}, \bibinfo{pages}{7192--7203} (\bibinfo{organization}{PMLR},
  \bibinfo{year}{2021}).

\bibitem{jin2018junction}
\bibinfo{author}{Jin, W.}, \bibinfo{author}{Barzilay, R.} \&
  \bibinfo{author}{Jaakkola, T.}
\newblock \bibinfo{title}{Junction tree variational autoencoder for molecular
  graph generation}.
\newblock In \emph{\bibinfo{booktitle}{International conference on machine
  learning}}, \bibinfo{pages}{2323--2332} (\bibinfo{organization}{PMLR},
  \bibinfo{year}{2018}).

\bibitem{jin2020hierarchical}
\bibinfo{author}{Jin, W.}, \bibinfo{author}{Barzilay, R.} \&
  \bibinfo{author}{Jaakkola, T.}
\newblock \bibinfo{title}{Hierarchical generation of molecular graphs using
  structural motifs}.
\newblock In \emph{\bibinfo{booktitle}{ICML}}, \bibinfo{pages}{4839--4848}
  (\bibinfo{organization}{PMLR}, \bibinfo{year}{2020}).

\bibitem{ganea2021geomol}
\bibinfo{author}{Ganea, O.} \emph{et~al.}
\newblock \bibinfo{title}{Geomol: Torsional geometric generation of molecular
  3d conformer ensembles}.
\newblock \emph{\bibinfo{journal}{Advances in Neural Information Processing
  Systems}} \textbf{\bibinfo{volume}{34}} (\bibinfo{year}{2021}).

\bibitem{xu2021end}
\bibinfo{author}{Xu, M.} \emph{et~al.}
\newblock \bibinfo{title}{An end-to-end framework for molecular conformation
  generation via bilevel programming}.
\newblock In \emph{\bibinfo{booktitle}{International Conference on Machine
  Learning}}, \bibinfo{pages}{11537--11547} (\bibinfo{organization}{PMLR},
  \bibinfo{year}{2021}).

\bibitem{shi2021learning}
\bibinfo{author}{Shi, C.}, \bibinfo{author}{Luo, S.}, \bibinfo{author}{Xu, M.}
  \& \bibinfo{author}{Tang, J.}
\newblock \bibinfo{title}{Learning gradient fields for molecular conformation
  generation}.
\newblock In \emph{\bibinfo{booktitle}{International Conference on Machine
  Learning}}, \bibinfo{pages}{9558--9568} (\bibinfo{organization}{PMLR},
  \bibinfo{year}{2021}).

\bibitem{liu2021pre}
\bibinfo{author}{Liu, S.} \emph{et~al.}
\newblock \bibinfo{title}{Pre-training molecular graph representation with 3d
  geometry}.
\newblock \emph{\bibinfo{journal}{International Conference on Learning
  Representations}}  (\bibinfo{year}{2022}).

\bibitem{mahmood2021masked}
\bibinfo{author}{Mahmood, O.}, \bibinfo{author}{Mansimov, E.},
  \bibinfo{author}{Bonneau, R.} \& \bibinfo{author}{Cho, K.}
\newblock \bibinfo{title}{Masked graph modeling for molecule generation}.
\newblock \emph{\bibinfo{journal}{Nature communications}}
  \textbf{\bibinfo{volume}{12}}, \bibinfo{pages}{1--12} (\bibinfo{year}{2021}).

\bibitem{hoffmann2019generating}
\bibinfo{author}{Hoffmann, M.} \& \bibinfo{author}{No{\'e}, F.}
\newblock \bibinfo{title}{Generating valid euclidean distance matrices}.
\newblock \emph{\bibinfo{journal}{arXiv preprint arXiv:1910.03131}}
  (\bibinfo{year}{2019}).

\bibitem{hoogeboom2022equivariant}
\bibinfo{author}{Hoogeboom, E.}, \bibinfo{author}{Satorras, V.~G.},
  \bibinfo{author}{Vignac, C.} \& \bibinfo{author}{Welling, M.}
\newblock \bibinfo{title}{Equivariant diffusion for molecule generation in 3d}.
\newblock \emph{\bibinfo{journal}{International Conference on Machine
  Learning}}  (\bibinfo{year}{2022}).

\bibitem{luo2021autoregressive}
\bibinfo{author}{Luo, Y.} \& \bibinfo{author}{Ji, S.}
\newblock \bibinfo{title}{An autoregressive flow model for 3d molecular
  geometry generation from scratch}.
\newblock In \emph{\bibinfo{booktitle}{International Conference on Learning
  Representations}} (\bibinfo{year}{2021}).

\bibitem{luo20213d}
\bibinfo{author}{Luo, S.}, \bibinfo{author}{Guan, J.}, \bibinfo{author}{Ma, J.}
  \& \bibinfo{author}{Peng, J.}
\newblock \bibinfo{title}{A 3d generative model for structure-based drug
  design}.
\newblock \emph{\bibinfo{journal}{Advances in Neural Information Processing
  Systems}} \textbf{\bibinfo{volume}{34}} (\bibinfo{year}{2021}).

\bibitem{mendez2021geometric}
\bibinfo{author}{M{\'e}ndez-Lucio, O.}, \bibinfo{author}{Ahmad, M.},
  \bibinfo{author}{del Rio-Chanona, E.~A.} \& \bibinfo{author}{Wegner, J.~K.}
\newblock \bibinfo{title}{A geometric deep learning approach to predict binding
  conformations of bioactive molecules}.
\newblock \emph{\bibinfo{journal}{Nature Machine Intelligence}}
  \textbf{\bibinfo{volume}{3}}, \bibinfo{pages}{1033--1039}
  (\bibinfo{year}{2021}).

\bibitem{liu2022generating}
\bibinfo{author}{Liu, M.}, \bibinfo{author}{Luo, Y.}, \bibinfo{author}{Uchino,
  K.}, \bibinfo{author}{Maruhashi, K.} \& \bibinfo{author}{Ji, S.}
\newblock \bibinfo{title}{Generating 3d molecules for target protein binding}.
\newblock \emph{\bibinfo{journal}{International Conference on Machine
  Learning}}  (\bibinfo{year}{2022}).

\bibitem{peng2022pocket2mol}
\bibinfo{author}{Peng, X.} \emph{et~al.}
\newblock \bibinfo{title}{Pocket2mol: Efficient molecular sampling based on 3d
  protein pockets}.
\newblock \emph{\bibinfo{journal}{International Conference on Machine
  Learning}}  (\bibinfo{year}{2022}).

\bibitem{liu2021spherical}
\bibinfo{author}{Liu, Y.} \emph{et~al.}
\newblock \bibinfo{title}{Spherical message passing for 3d graph networks}.
\newblock \emph{\bibinfo{journal}{International Conference on Learning
  Representations}}  (\bibinfo{year}{2022}).

\bibitem{satorras2021n}
\bibinfo{author}{Satorras, V.~G.}, \bibinfo{author}{Hoogeboom, E.},
  \bibinfo{author}{Fuchs, F.~B.}, \bibinfo{author}{Posner, I.} \&
  \bibinfo{author}{Welling, M.}
\newblock \bibinfo{title}{E (n) equivariant normalizing flows}.
\newblock \emph{\bibinfo{journal}{NeurIPS}}  (\bibinfo{year}{2021}).

\bibitem{papamakarios2021normalizing}
\bibinfo{author}{Papamakarios, G.}, \bibinfo{author}{Nalisnick, E.},
  \bibinfo{author}{Rezende, D.~J.}, \bibinfo{author}{Mohamed, S.} \&
  \bibinfo{author}{Lakshminarayanan, B.}
\newblock \bibinfo{title}{Normalizing flows for probabilistic modeling and
  inference}.
\newblock \emph{\bibinfo{journal}{Journal of Machine Learning Research}}
  \textbf{\bibinfo{volume}{22}}, \bibinfo{pages}{1--64} (\bibinfo{year}{2021}).

\bibitem{dinh2014nice}
\bibinfo{author}{Dinh, L.}, \bibinfo{author}{Krueger, D.} \&
  \bibinfo{author}{Bengio, Y.}
\newblock \bibinfo{title}{Nice: Non-linear independent components estimation}.
\newblock \emph{\bibinfo{journal}{arXiv preprint arXiv:1410.8516}}
  (\bibinfo{year}{2014}).

\bibitem{dinh2016density}
\bibinfo{author}{Dinh, L.}, \bibinfo{author}{Sohl-Dickstein, J.} \&
  \bibinfo{author}{Bengio, S.}
\newblock \bibinfo{title}{Density estimation using real nvp}.
\newblock \emph{\bibinfo{journal}{arXiv preprint arXiv:1605.08803}}
  (\bibinfo{year}{2016}).

\bibitem{o2011open}
\bibinfo{author}{O'Boyle, N.~M.} \emph{et~al.}
\newblock \bibinfo{title}{Open babel: An open chemical toolbox}.
\newblock \emph{\bibinfo{journal}{Journal of cheminformatics}}
  \textbf{\bibinfo{volume}{3}}, \bibinfo{pages}{1--14} (\bibinfo{year}{2011}).

\bibitem{kim2015universal}
\bibinfo{author}{Kim, Y.} \& \bibinfo{author}{Kim, W.~Y.}
\newblock \bibinfo{title}{Universal structure conversion method for organic
  molecules: from atomic connectivity to three-dimensional geometry}.
\newblock \emph{\bibinfo{journal}{Bulletin of the Korean Chemical Society}}
  \textbf{\bibinfo{volume}{36}}, \bibinfo{pages}{1769--1777}
  (\bibinfo{year}{2015}).

\bibitem{ramakrishnan2014quantum}
\bibinfo{author}{Ramakrishnan, R.}, \bibinfo{author}{Dral, P.~O.},
  \bibinfo{author}{Rupp, M.} \& \bibinfo{author}{Von~Lilienfeld, O.~A.}
\newblock \bibinfo{title}{Quantum chemistry structures and properties of 134
  kilo molecules}.
\newblock \emph{\bibinfo{journal}{Scientific data}}
  \textbf{\bibinfo{volume}{1}}, \bibinfo{pages}{1--7} (\bibinfo{year}{2014}).

\bibitem{garcia2021n}
\bibinfo{author}{Garcia~Satorras, V.}, \bibinfo{author}{Hoogeboom, E.},
  \bibinfo{author}{Fuchs, F.}, \bibinfo{author}{Posner, I.} \&
  \bibinfo{author}{Welling, M.}
\newblock \bibinfo{title}{E (n) equivariant normalizing flows}.
\newblock \emph{\bibinfo{journal}{Advances in Neural Information Processing
  Systems}} \textbf{\bibinfo{volume}{34}} (\bibinfo{year}{2021}).

\bibitem{sun2018pyscf}
\bibinfo{author}{Sun, Q.} \emph{et~al.}
\newblock \bibinfo{title}{Pyscf: the python-based simulations of chemistry
  framework}.
\newblock \emph{\bibinfo{journal}{Wiley Interdisciplinary Reviews:
  Computational Molecular Science}} \textbf{\bibinfo{volume}{8}},
  \bibinfo{pages}{e1340} (\bibinfo{year}{2018}).

\bibitem{sun2020recent}
\bibinfo{author}{Sun, Q.} \emph{et~al.}
\newblock \bibinfo{title}{Recent developments in the pyscf program package}.
\newblock \emph{\bibinfo{journal}{The Journal of chemical physics}}
  \textbf{\bibinfo{volume}{153}}, \bibinfo{pages}{024109}
  (\bibinfo{year}{2020}).

\bibitem{anderson2003process}
\bibinfo{author}{Anderson, A.~C.}
\newblock \bibinfo{title}{The process of structure-based drug design}.
\newblock \emph{\bibinfo{journal}{Chemistry \& biology}}
  \textbf{\bibinfo{volume}{10}}, \bibinfo{pages}{787--797}
  (\bibinfo{year}{2003}).

\bibitem{tripathi2017molecular}
\bibinfo{author}{Tripathi, A.} \& \bibinfo{author}{Bankaitis, V.~A.}
\newblock \bibinfo{title}{Molecular docking: from lock and key to combination
  lock}.
\newblock \emph{\bibinfo{journal}{Journal of molecular medicine and clinical
  applications}} \textbf{\bibinfo{volume}{2}} (\bibinfo{year}{2017}).

\bibitem{alon2021structures}
\bibinfo{author}{Alon, A.} \emph{et~al.}
\newblock \bibinfo{title}{Structures of the $\sigma$2 receptor enable docking
  for bioactive ligand discovery}.
\newblock \emph{\bibinfo{journal}{Nature}} \textbf{\bibinfo{volume}{600}},
  \bibinfo{pages}{759--764} (\bibinfo{year}{2021}).

\bibitem{francoeur2020three}
\bibinfo{author}{Francoeur, P.~G.} \emph{et~al.}
\newblock \bibinfo{title}{Three-dimensional convolutional neural networks and a
  cross-docked data set for structure-based drug design}.
\newblock \emph{\bibinfo{journal}{Journal of chemical information and
  modeling}} \textbf{\bibinfo{volume}{60}}, \bibinfo{pages}{4200--4215}
  (\bibinfo{year}{2020}).

\bibitem{trott2010autodock}
\bibinfo{author}{Trott, O.} \& \bibinfo{author}{Olson, A.~J.}
\newblock \bibinfo{title}{Autodock vina: improving the speed and accuracy of
  docking with a new scoring function, efficient optimization, and
  multithreading}.
\newblock \emph{\bibinfo{journal}{Journal of computational chemistry}}
  \textbf{\bibinfo{volume}{31}}, \bibinfo{pages}{455--461}
  (\bibinfo{year}{2010}).

\bibitem{alhossary2015fast}
\bibinfo{author}{Alhossary, A.}, \bibinfo{author}{Handoko, S.~D.},
  \bibinfo{author}{Mu, Y.} \& \bibinfo{author}{Kwoh, C.-K.}
\newblock \bibinfo{title}{Fast, accurate, and reliable molecular docking with
  quickvina 2}.
\newblock \emph{\bibinfo{journal}{Bioinformatics}}
  \textbf{\bibinfo{volume}{31}}, \bibinfo{pages}{2214--2216}
  (\bibinfo{year}{2015}).

\bibitem{bento2020open}
\bibinfo{author}{Bento, A.~P.} \emph{et~al.}
\newblock \bibinfo{title}{An open source chemical structure curation pipeline
  using rdkit}.
\newblock \emph{\bibinfo{journal}{Journal of Cheminformatics}}
  \textbf{\bibinfo{volume}{12}}, \bibinfo{pages}{1--16} (\bibinfo{year}{2020}).

\bibitem{rappe1992uff}
\bibinfo{author}{Rapp{\'e}, A.~K.}, \bibinfo{author}{Casewit, C.~J.},
  \bibinfo{author}{Colwell, K.}, \bibinfo{author}{Goddard~III, W.~A.} \&
  \bibinfo{author}{Skiff, W.~M.}
\newblock \bibinfo{title}{Uff, a full periodic table force field for molecular
  mechanics and molecular dynamics simulations}.
\newblock \emph{\bibinfo{journal}{Journal of the American chemical society}}
  \textbf{\bibinfo{volume}{114}}, \bibinfo{pages}{10024--10035}
  (\bibinfo{year}{1992}).

\bibitem{ragoza2022generating}
\bibinfo{author}{Ragoza, M.}, \bibinfo{author}{Masuda, T.} \&
  \bibinfo{author}{Koes, D.~R.}
\newblock \bibinfo{title}{Generating 3d molecules conditional on receptor
  binding sites with deep generative models}.
\newblock \emph{\bibinfo{journal}{Chemical science}}
  \textbf{\bibinfo{volume}{13}}, \bibinfo{pages}{2701--2713}
  (\bibinfo{year}{2022}).

\bibitem{zhang2022hierarchical}
\bibinfo{author}{Zhang, Z.}, \bibinfo{author}{Liu, Q.}, \bibinfo{author}{Hu,
  Q.} \& \bibinfo{author}{Lee, C.-K.}
\newblock \bibinfo{title}{Hierarchical graph transformer with adaptive node
  sampling}.
\newblock \emph{\bibinfo{journal}{Advances in Neural Information Processing
  Systems}}  (\bibinfo{year}{2022}).

\bibitem{zhang2021graphmi}
\bibinfo{author}{Zhang, Z.} \emph{et~al.}
\newblock \bibinfo{title}{Graphmi: Extracting private graph data from graph
  neural networks}.
\newblock \emph{\bibinfo{journal}{IJCAI}}  (\bibinfo{year}{2021}).

\bibitem{zhang2022model}
\bibinfo{author}{Zhang, Z.} \emph{et~al.}
\newblock \bibinfo{title}{Model inversion attacks against graph neural
  networks}.
\newblock \emph{\bibinfo{journal}{IEEE Transactions on Knowledge and Data
  Engineering}}  (\bibinfo{year}{2022}).

\bibitem{zhang2021backdoor}
\bibinfo{author}{Zhang, Z.}, \bibinfo{author}{Jia, J.}, \bibinfo{author}{Wang,
  B.} \& \bibinfo{author}{Gong, N.~Z.}
\newblock \bibinfo{title}{Backdoor attacks to graph neural networks}.
\newblock In \emph{\bibinfo{booktitle}{Proceedings of the 26th ACM Symposium on
  Access Control Models and Technologies}}, \bibinfo{pages}{15--26}
  (\bibinfo{year}{2021}).

\bibitem{zhang2022protgnn}
\bibinfo{author}{Zhang, Z.}, \bibinfo{author}{Liu, Q.}, \bibinfo{author}{Wang,
  H.}, \bibinfo{author}{Lu, C.} \& \bibinfo{author}{Lee, C.}
\newblock \bibinfo{title}{Protgnn: Towards self-explaining graph neural
  networks}.
\newblock In \emph{\bibinfo{booktitle}{Proceedings of the AAAI Conference on
  Artificial Intelligence}}, vol.~\bibinfo{volume}{36},
  \bibinfo{pages}{9127--9135} (\bibinfo{year}{2022}).

\bibitem{steinegger2017mmseqs2}
\bibinfo{author}{Steinegger, M.} \& \bibinfo{author}{S{\"o}ding, J.}
\newblock \bibinfo{title}{Mmseqs2 enables sensitive protein sequence searching
  for the analysis of massive data sets}.
\newblock \emph{\bibinfo{journal}{Nature biotechnology}}
  \textbf{\bibinfo{volume}{35}}, \bibinfo{pages}{1026--1028}
  (\bibinfo{year}{2017}).

\bibitem{simm2020reinforcement}
\bibinfo{author}{Simm, G.}, \bibinfo{author}{Pinsler, R.} \&
  \bibinfo{author}{Hern{\'a}ndez-Lobato, J.~M.}
\newblock \bibinfo{title}{Reinforcement learning for molecular design guided by
  quantum mechanics}.
\newblock In \emph{\bibinfo{booktitle}{International Conference on Machine
  Learning}}, \bibinfo{pages}{8959--8969} (\bibinfo{organization}{PMLR},
  \bibinfo{year}{2020}).

\bibitem{vaswani2017attention}
\bibinfo{author}{Vaswani, A.} \emph{et~al.}
\newblock \bibinfo{title}{Attention is all you need}.
\newblock \emph{\bibinfo{journal}{Advances in neural information processing
  systems}} \textbf{\bibinfo{volume}{30}} (\bibinfo{year}{2017}).

\bibitem{kingma2014adam}
\bibinfo{author}{Kingma, D.~P.} \& \bibinfo{author}{Ba, J.}
\newblock \bibinfo{title}{Adam: A method for stochastic optimization}.
\newblock \emph{\bibinfo{journal}{arXiv preprint arXiv:1412.6980}}
  (\bibinfo{year}{2014}).

\end{thebibliography}

\section*{Acknowledgements}
This research was partially supported by grants from the National Natural Science Foundation of China (Grants No.61922073 and U20A20229).

\section*{Author contributions statement}
Z.X.Z, Q.L, C.L., C.H., and E.H.C. designed the research,  Z.X.Z conducted the experiments, Z.X.Z, Q.L, and C.L. analyzed the results.  All authors reviewed the manuscript. 

\section*{Competing interests}
The authors declare no competing interests.

\section*{Additional information}

{\bf Correspondence and requests for material} should be addressed to Qi Liu.

\clearpage
\setcounter{figure}{0}
\setcounter{table}{0}
\makeatletter 
\renewcommand{\thefigure}{S\@arabic\c@figure}
\renewcommand{\thetable}{S\@arabic\c@table}
\makeatother

\section{Supplementary Information}
\begin{table}[h]
\caption{{\bf Substructure analysis of the generated molecules.} The KL divergence of the bond lengths (upper part) and bond angles (lower part) between the training set and the generated molecules are shown below.}
\label{substructure}
\centering
\begin{tabular}{c|cccccc}
\toprule
    Distances/Angles & E-NFs & G-SchNet & G-SphereNet& EDM & MolCode (w/o bond) & MolCode  \\ \midrule
    CC&0.53&0.44 &0.30 &0.36 &0.32 &{\bf0.24}\\
    CN&0.87&0.68&0.45&0.37&0.43&{\bf0.30}\\
    CO&0.49&0.32&0.24&0.26&0.25&{\bf0.21}\\
    NO&0.39&0.27&0.20&0.24&0.19&{\bf0.17}\\
    \midrule
    CCC & 1.25  & 0.96  &0.65& 0.48 &0.66 &\textbf{0.25}\\
    CCO & 0.98  & 0.85  &0.41& 0.33 &0.47 &\textbf{0.23}  \\ 
    CNC & 1.44 & 1.37  &0.71& 0.56 &0.64 &\textbf{0.42}  \\ 
    CCN & 1.30  & 0.95  &0.74& 0.84 &0.62 &\textbf{0.37}  \\  
    \bottomrule
\end{tabular}
\end{table}
\newpage
\begin{figure*}[h]
	\centering
	\subfigure[$z_i^V$]{\includegraphics[width=0.32\linewidth]{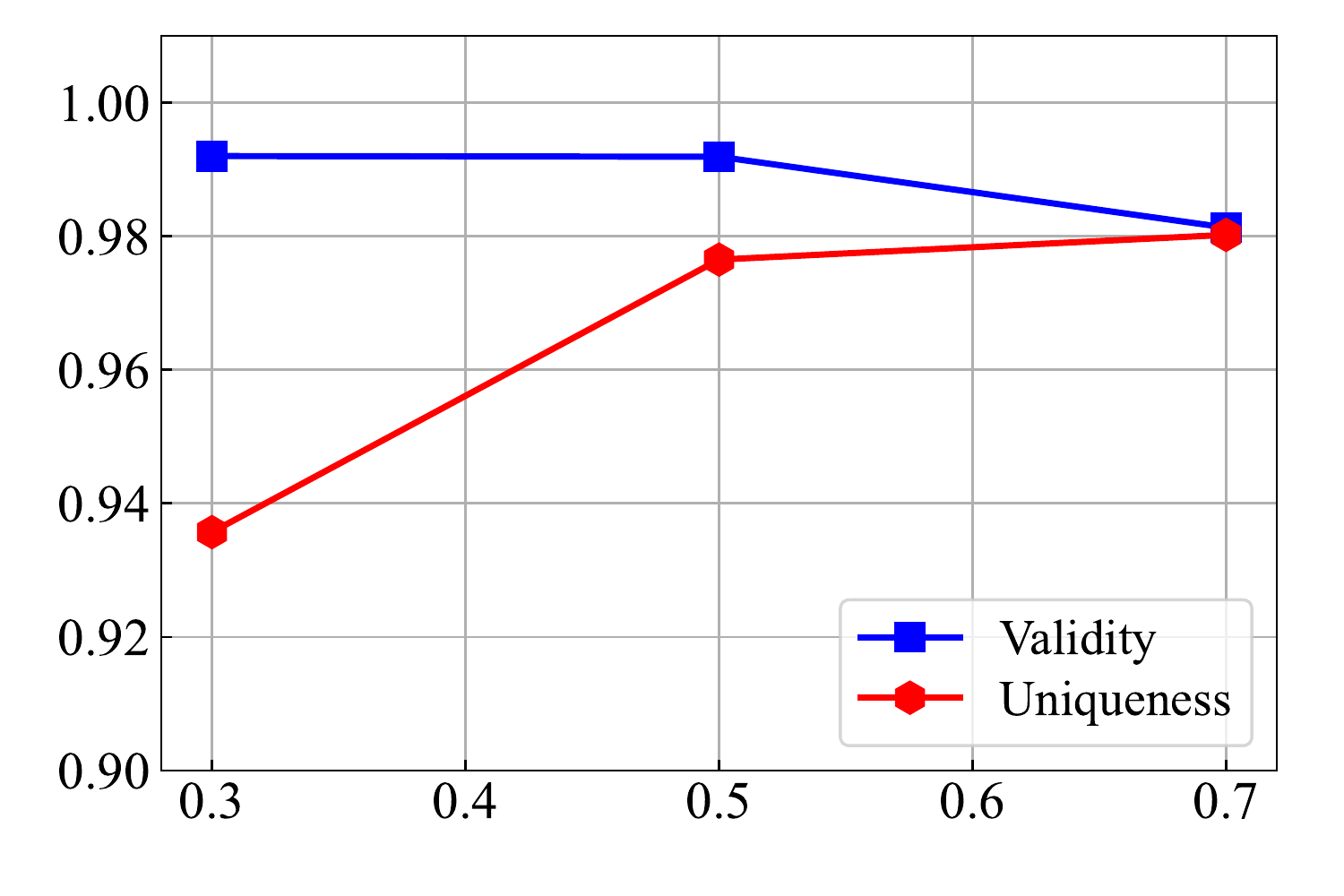}}
    \subfigure[$z_i^A$]{\includegraphics[width=0.32\linewidth]{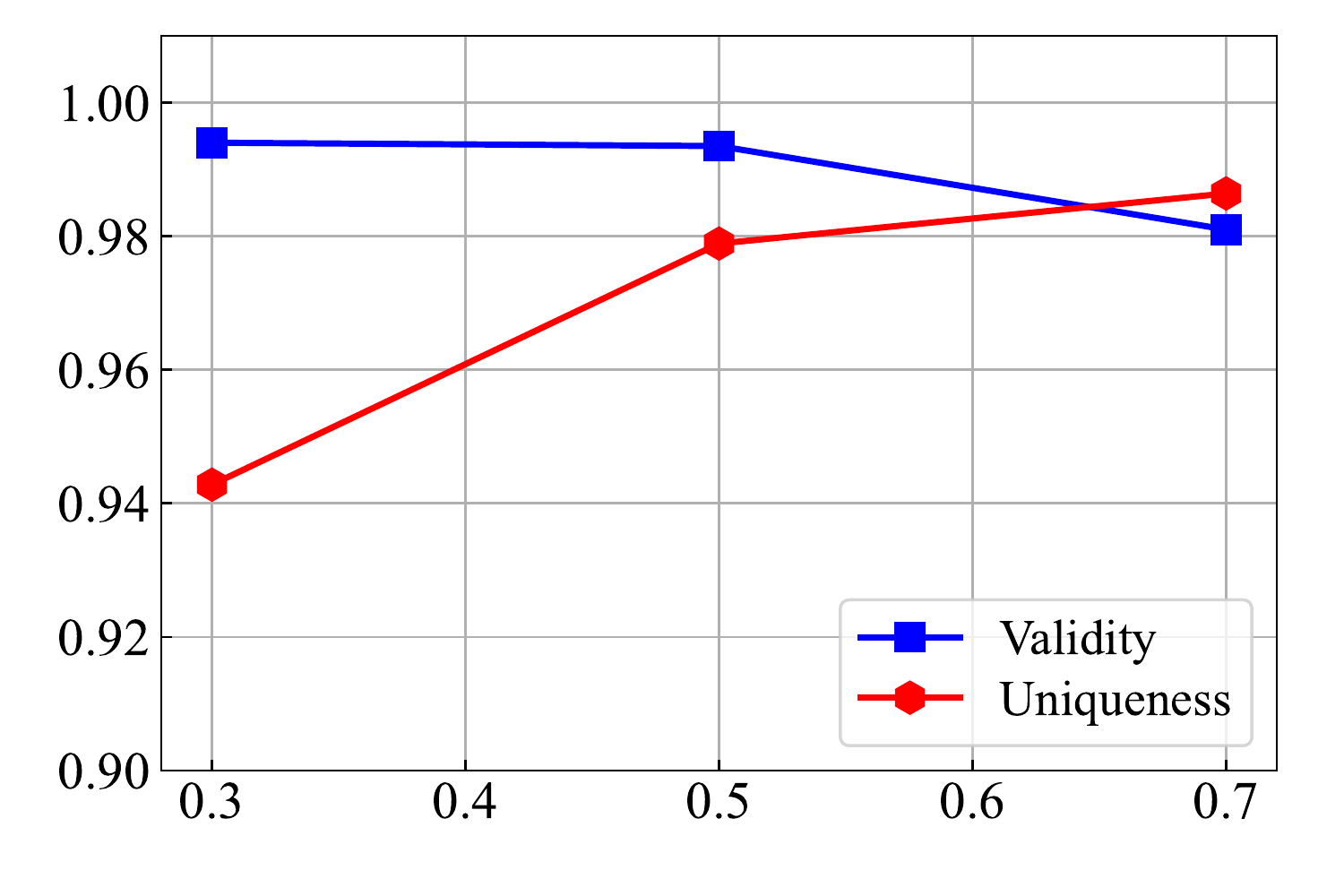}}
    \subfigure[$z_i^d$]{\includegraphics[width=0.32\linewidth]{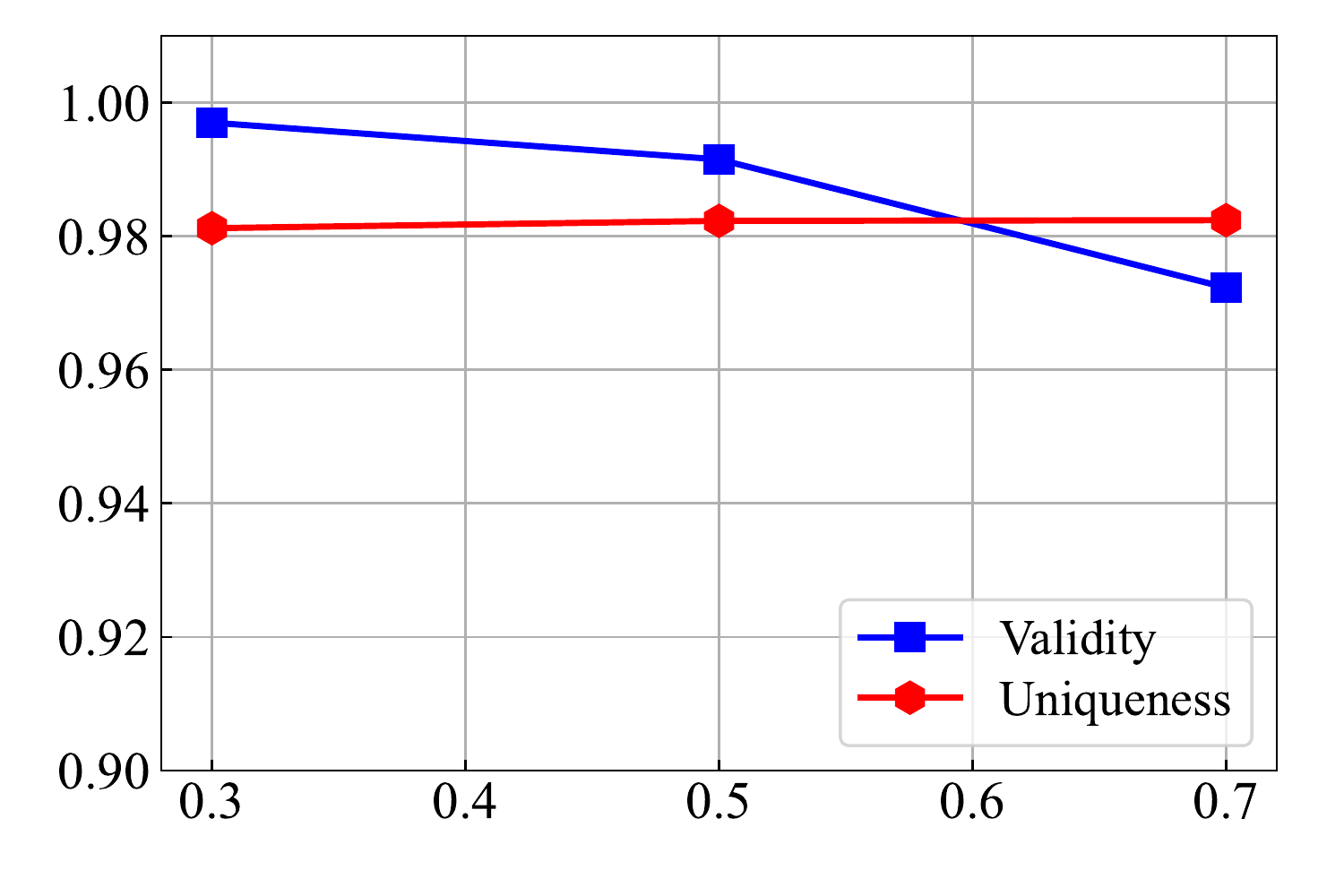}}\\
    \subfigure[$z_i^\theta$]{\includegraphics[width=0.32\linewidth]{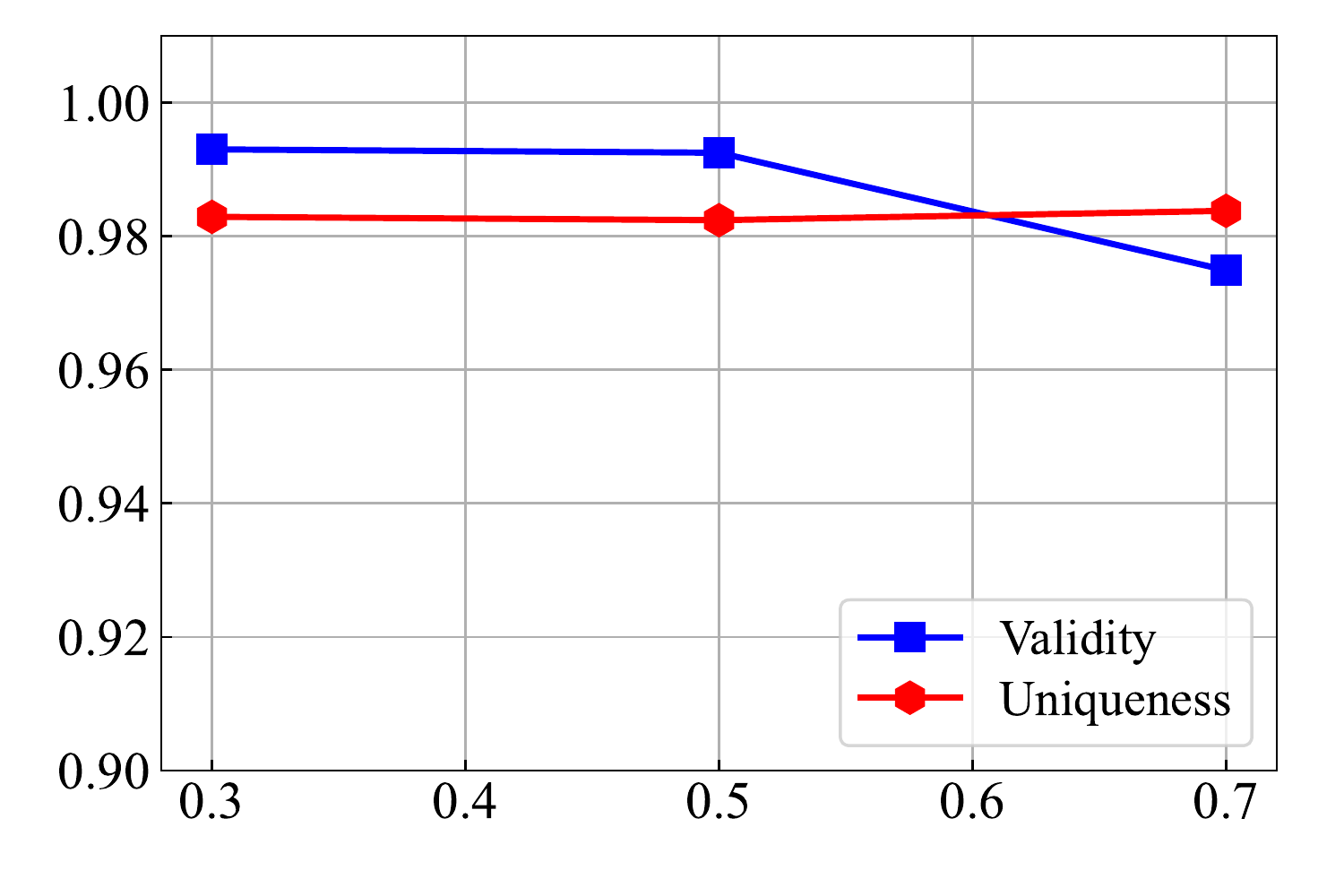}}
    \subfigure[$z_i^\phi$]{\includegraphics[width=0.32\linewidth]{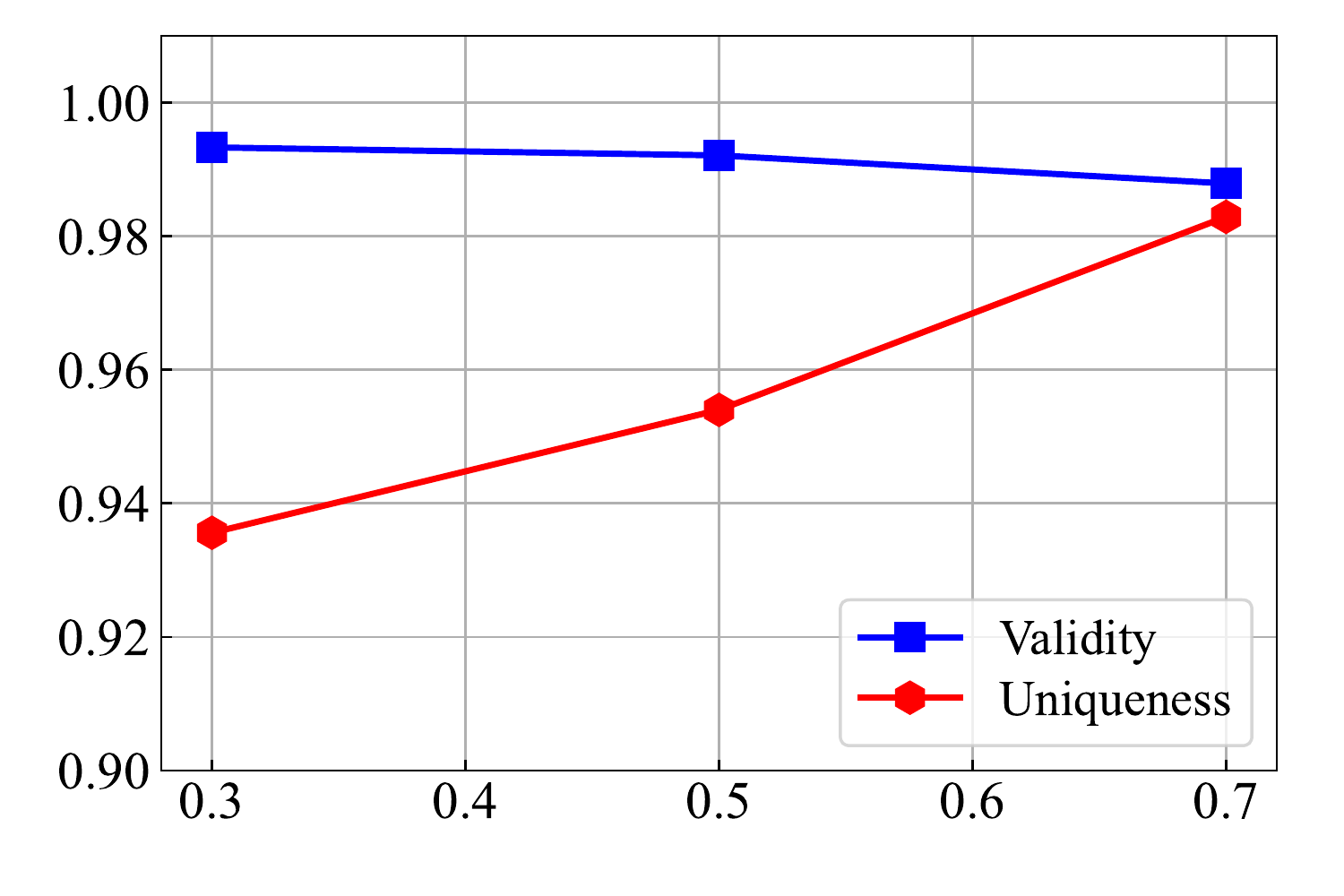}}
	\caption{Influence of temperature hyperparameters on Validity and Uniqueness in the random molecule generation task.}
	\label{app:hyper}
\end{figure*}

\newpage

\begin{algorithm}[!h]
\caption{Training Algorithm of MolCode}
\label{alg:train}
\textbf{Input}: Molecular dataset $\mathcal{M}$, learning rate $\eta$, Adam hyperparameters $\beta_1, \beta_2$, batch size $B$, GoGen model with trainable parameter $w$, latent distribution $p_{Z_V}, p_{Z_A}, p_{Z_d}, p_{Z_\theta}, p_{Z_\phi}$, maximum number of atoms $n$

\textbf{Initial}: Parameters $w$ of MolCode
\begin{algorithmic}[1] 
\WHILE{$w$ is not converged}
    \STATE Sample a batch of $B$ molecule $mol$ from dataset $\mathcal{M}$
    \STATE $L=0$
    \FOR{$G\in mol$}
    \STATE Set $n$ as the number of atoms in $G$ and order the atoms in $G$
    \FOR{$i=1,...,n-1$}
        \STATE Get $V_i, d_i, \theta_i$ (if $i \ge 2$), $\phi_i$ (if $i \ge 3$) and the reference atoms $(f,c,e)$
        \STATE Get $z_i^V, z_i^d, z_i^\theta$ (if $i \ge 2$), $z_i^\phi$ (if $i \ge 3$) with the flow modules in MolCode
        \STATE $L = L-log p_{Z_V} (z_i^V)-log p_{Z_d} (z_i^d)$
        \STATE $L = L-log p_{Z_V} (z_i^\theta)$ (if $i \ge 2$)
        \STATE $L = L-log p_{Z_V} (z_i^\phi)$ (if $i \ge 3$)
        \FOR{$j\in \{f,c,e\}$}
            \STATE Get $A_{ij}$ and $z_{ij}^A$
            \STATE $L = L-log p_{Z_A} (z_{ij}^A)$
        \ENDFOR
        \STATE Add the binary cross entropy loss for the focal atom selection to $L$
    \ENDFOR
    \ENDFOR
    \STATE $w \leftarrow \text{ADAM} (\frac{L}{B}, w, \eta, \beta_1, \beta_2)$
\ENDWHILE
\end{algorithmic}
\end{algorithm}

\begin{algorithm}[!h]
\caption{Generation Algorithm of MolCode}
\label{alg:generate}
\textbf{Input}: GoGen  model with parameter $w$, latent distribution $p_{Z_V}, p_{Z_A}, p_{Z_d}, p_{Z_\theta}, p_{Z_\phi}$, maximum number of atoms $n$, maximum number of trials to sample bond types $T$
\begin{algorithmic}[1] 
\FOR{$i=1,...,n-1$}
    \STATE Initialize molecular graph $G_1$ with one carbon atom, whose coordinate is $R_0$ = [0, 0, 0]
    \STATE Sample $z_i^V \sim p_{Z_V}$ and generate $V_i$
    \STATE Get the candidate focal atom set by the atom-wise classifier
    \STATE Get the reference atoms $\{f,c,e\}$
    \FOR{$j\in \{f,c,e\}$}
    \STATE Count = 0
            \STATE Get $z_{ij}^A \sim p_{Z_A}$ and generate $A_{ij}$  
            \STATE{$\textbf{if}$ $\sum_j |A_{ij}| \ge {\rm Valency}(X_i) {~\rm or~} \sum_i |A_{ij}| \ge {\rm Valency}(X_j)$ and Count $\le T$} $\textbf{then}$\par
            $\indent$ {Reject $A_{ij}$ and sample a new $z_{ij}^A$; Count+=1}
            \STATE $\textbf{else}$ \par $\indent$ Assign no bond to $A_{ij}$
            \STATE $\textbf{end if}$
    \ENDFOR
    \STATE {$\textbf{if}$ the candidate focal atom set is empty or $\sum_j |A_{ij}| = 0$} $\textbf{then}$
        \STATE $\indent$ Output $G_i$
    \STATE $\textbf{else}$ \par
        $\indent$ Randomly select the focal atom $f$ from the candidate focal atom set
    \STATE $\textbf{end if}$
    \STATE Sample $z_i^d, z_i^\theta$ (if $i \ge 2$), $z_i^\phi$ (if $i \ge 3$)
    \STATE Generate $d_i, \theta_i$ (if $i \ge 2$), $\phi_i$ (if $i \ge 3$) and get $R_i$, update $G_i$ to $G_{i+1}$
    \ENDFOR
\STATE Output $G_n$
\end{algorithmic}
\end{algorithm}

\end{document}